\documentclass[journal]{IEEEtran}
\usepackage[T1]{fontenc}
\ifCLASSINFOpdf
   \usepackage[pdftex]{graphicx}
\else
   \usepackage[dvips]{graphicx}
\fi
\usepackage{multirow}
\usepackage{url}
\usepackage{amssymb}

\usepackage[cmex10]{amsmath}
\DeclareMathOperator*{\argmax}{\arg\!\max}

\usepackage[normalem]{ulem}
\begin{document}
%
%

\hyphenation{op-tical net-works semi-conduc-tor}

%
\title{In-Band-Interference Robust Synchronization Algorithm for an NC-OFDM System}
%
%
%

\author{Pawel~Kryszkiewicz,~\IEEEmembership{Student Member,~IEEE,}
        and~Hanna~Bogucka,~\IEEEmembership{Senior Member,~IEEE}
\thanks{The authors are with the Chair of Wireless Communications, Poznan University of Technology, Poznan, Poland, e-mail: pawel.kryszkiewicz@put.poznan.pl.
}
\thanks {
The work presented in this paper has been funded by the National Science Centre, Poland, based on decision no. DEC-2012/05/N/ST7/00164 for PRELUDIUM project and by the Polish Ministry of Science and Higher Education within the status activity task \emph{Cognitive radio systems} in 2016.}
\thanks{Copyright (c) 2016 IEEE. Personal use is permitted. For any other purposes, permission must be obtained from the IEEE by emailing pubs-permissions@ieee.org.
This is the author's version of an article that has been published in this journal. Changes were made to this version by the publisher prior to publication.
The final version of record is available at http://dx.doi.org/10.1109/TCOMM.2016.2540640}
}

%
%

\markboth{IEEE TRANSACTIONS ON COMMUNICATIONS}%
{SUBMITTED PAPER}
%



\maketitle

\begin{abstract}
We consider receiver synchronization in the Non-Contiguous Orthogonal Frequency Division Multiplexing (NC-OFDM) -based radio system in the presence of in-band interfering signal, which occupies the frequency-band between blocks of subcarriers (SCs) used by this system, i.e. in-band of NC-OFDM receiver spectrum range. The paper proposes a novel preamble-based synchronization algorithm for estimation of the time and frequency offset based on the received signal cross-correlation with the reference preamble. Contrary to the existing algorithms, it is robust against in-band interference including narrowband interference at the cost of increased complexity. Moreover, in the interference-free system, the probability of frame synchronization error is improved in comparison to all simulated algorithms.
\end{abstract}


%

\section{Introduction}
%
%
%
%
\IEEEPARstart{T}{he} Non-Contiguous Orthogonal Frequency Division Multiplexing (NC-OFDM) is considered as a flexible transmission technique able to aggregate  fragmented frequency bands by modulating only some, possibly non-contiguous OFDM subcarriers \cite{Bogucka_Radar_Sonar_2011}. It is a suitable transmission technique for cognitive radio (CR) to make an opportunistic use of spectrum resources \cite{Mahmoud09_OFDM_for_CR}. A CR system is required to protect the licensed-user (LU) transmission by keeping the created interference power observed at the LU receiver below the required level, e.g. by applying the guard subcarriers (GSs) on the edges of a CR NC-OFDM system useful bands (neighboring the LU's band), time-domain windowing \cite{Weiss_oknowanie_Re_i_TX} or some more sophisticated methods \cite{KryszkiewiczEURASIP12,Kryszkiewicz_OCCS_2013}.  
However, an LU system is not required to take any measures for the protection of the CR transmission.

The CR NC-OFDM signal-detection algorithm can take advantage of the fact that an LU uses the frequencies unused by the CR inside its band, and suppress the interference by the means of the frequency-domain data symbols detection using FFT block (with addition of time-domain windowing for extra interference reduction) \cite{Weiss_oknowanie_Re_i_TX}.
However, it requires precise prior synchronization of the NC-OFDM frame in the time- and in the frequency domain which is recognized as a challenge when the  in-band interference is present \cite{Brandes_mitigation_NB_interf}.

In the literature, synchronization algorithms developed for OFDM (overviewed in \cite{Morelli_07_overview_synchr}) have been also considered for NC-OFDM. The preamble-based Schmidl\&Cox (S\&C) algorithm \cite{Schmidl_Cox_1997} is the most known. There, the preamble preceding data symbols consists of two identical series of time-samples. We will refer to it as the \emph{S\&C preamble}. The receiver determines auto-correlation in the time domain to find the \emph{optimal timing point}, i.e. the first effective sample of the first OFDM symbol in a frame (after omission of the Cyclic Prefix (CP)). This autocorrelation is also used for the estimation of fractional carrier-frequency offset (CFO)
normalized to the OFDM subcarriers (SCs) spacing. 
The S\&C algorithm has been improved e.g. in \cite{Minn_2003_improved_S_C} and \cite{Morelli_freq_improv_S_C}.

In our considered case of a CR system, performance deterioration can be caused by the presence of high power LU-originating interference. Recently, an S\&C algorithm has been evaluated for an OFDM system operating in the presence of various interference types: narrowband modeled as a complex sinusoid \cite{Coulson_2004_synchr_narrow_sinusoid}, narrowband digitally-modulated signal \cite{Marey_2007_SC_narrowband} and wideband occupying all SCs adjacent to the block of SCs used by a CR system \cite{Zivkovic_2011_synchr_SC_wideband}. 
The wideband interference degrades
the synchronization performance as if the preamble was distorted only by the Additive White Gaussian Noise (AWGN) of the same power as the interference. 
Even more difficult is NC-OFDM synchronization in case of narrowband interference (NBI). 
As shown in \cite{Coulson_2004_synchr_narrow_sinusoid}, in the absence of a useful signal, the synchronization algorithm can result in the \emph{false synchronization} (interference can be detected as a useful signal). This is caused by the auto-correlation properties of the complex sinusoid. As the bandwidth of the interfering signal increases, this effect is suppressed \cite{Marey_2007_SC_narrowband}. 
In \cite{Awoseyila_synchronization}, the S\&C algorithm is used to obtain coarse time- and fractional frequency synchronization. Afterward, cross-correlation is used to localize the first channel-path component, and to estimate the integer CFO. Unfortunately, this framework is not useful in the presence of in-band interference, as it is very probable that the first step (auto-correlation based) of the algorithm will fail preventing the success of the second stage.

One approach to the NBI rejection at the NC-OFDM receiver is to suppress it by filtering. As shown in \cite{Coulson_2004_synchr_narrow_sinusoid}, the a priori knowledge or correct detection of the interference center-frequency is required for this purpose.
High reliability of frame detection can be obtained by the frequency-domain filtering and cross-correlation \cite{Sun_2010_synchr_filter}, assuming idealistic channel with no fading and no CFO. Finally, filtering has major practical issues. The filter has to be adjusted to the interference characteristic, and redesigned as the CFO or an interfering signal frequency change. Moreover, low-order filters can distort the useful signal \cite{Brandes_mitigation_NB_interf} while high-order filters have high complexity, and contribute to the time-domain signal dispersion \cite{Faulkner_filtering_effect_OFDM}. Another possibility is to use the autocorrelation-based method robust against NBI as discussed in \cite{Sanguinetti_2010_synchr}. Although, it might be optimal for OFDM, it does not take advantage of the fact that the interference band and the NC-OFDM-utilized subcarriers are non-overlapping. Moreover, it has a limited range of CFO estimation, and because the preamble is multiplied in the time-domain by a bipolar sequence, the signal spectrum changes, which may be a source of interference generated to the LU band.  

The utilization of cross-correlation between the received preamble and the reference (original) preamble can be more advantageous 
than the autocorrelation-based methods in the presence of NBI. Combination of cross-correlation-based synchronization with spectrum sensing has been proposed in \cite{Saha_Dyspan_synchronization}, however, no measures against the multipath effects or the non-zero CFO have been proposed.
The algorithm proposed in \cite{Ziabari_synchr_NC_OFDM} uses modified cross-correlation of adjustable length to obtain coarse time synchronization. Fine time- and the integer-CFO estimation is obtained as in \cite{Awoseyila_synchronization}. Additionally, a fractional-CFO estimator uses the cross-correlation amplitude of the strongest path only, as in \cite{Ren_freq_estimation}, resulting in limited accuracy. The main drawbacks of this algorithm are high computational complexity and limited robustness to CFO.

In this paper, we present a new synchronization algorithm for an NC-OFDM CR system, robust against an LU-originating interference. We assume that the LU transmission is protected from the CR-generated interference 
by appropriate spectrum managing, e.g., by the CR ocuppying only subcarriers unused by LUs, and spectrum
shaping using the GSs, windowing or other spectrum shaping methods \cite{KryszkiewiczEURASIP12}. 
We call our algorithm \emph{Licensed-User Insensitive Synchronization Algorithm} (LUISA). It is based on cross-correlation of the received and reference preamble (RP) available at the receiver. The algorithm is insensitive to interference non-overlapping, in the frequency domain, the utilized NC-OFDM subcarriers. Even in the no-interference scenario, the probability of synchronization error is lower than in \cite{Schmidl_Cox_1997}, \cite{Sanguinetti_2010_synchr} and \cite{Ziabari_synchr_NC_OFDM}. 
It makes use of all signal path-components with their phase dependencies, which allows for the decrease of the carrier-offset and timing-offset estimation MSE in comparison to \cite{Ziabari_synchr_NC_OFDM}. 

In Section II, we present the NC-OFDM system model. In Section III, our new proposed synchronization algorithm is presented in steps. In Section IV, simulation results are presented and computational complexity of the algorithm is evaluated. Section V presents conclusions.

\section{System model}
The NC-OFDM based CR transmitter uses $N$-point Inverse Fast Fourier Transform (IFFT) for multicarrier modulation. In each frame, $P$ NC-OFDM symbols are transmitted, and each OFDM symbol is preceded by a CP of $N_{\mathrm{CP}}$ samples. 
Let us denote a vector of complex symbols at the input of the IFFT as $\mathbf{d}^{(p)}=\left\{ d_{k}^{(p)}\right\}$, where $k=-N/2,...,N/2-1$ is the SC index, and $p=0,...,P-1$ is the OFDM symbol index. Note that $\alpha$ out of $N$ SCs are modulated by complex symbols. The indices of these SCs are elements of vector $\mathbf{I}=\{I_{c}\}$ for $c=1,...,\alpha$ and $I_{c}\in \{-N/2,...,N/2-1\}$. The other SCs (including GSs at the band edges) are modulated by zeros
in order to protect the LU transmission and to allow digital-analog conversion, i.e. $d_{k}^{(p)}=0$ for $k\in\{-N/2,...,N/2-1\} \setminus \mathbf{I}$. The non-contiguity of the considered NC-OFDM system SCs results from the CR scenario-dependent choice of SCs indices $I_{c}$, i.e. it is  not a series of consecutive integer numbers as in a typical OFDM system. The indices of subcarriers modulated by the preamble symbols are elements of vector $\mathbf{I_{\mathrm{RP}}}$ (where $\mathbf{I_{\mathrm{RP}}}\subseteq \mathbf{I}$), and depend on the preamble type as shown in Sec. \ref{sec_statistical_properties}. The $n$-th sample of the $p$-th OFDM symbol at the output of IFFT and after the CP addition is defined as: 
\begin{equation}
x^{(p)}_{n}=\left\{
\begin{array}{ll}
\frac{1}{\sqrt{N}}\sum_{k=-N/2}^{N/2-1}d_{k}^{(p)}e^{j2\pi \frac{nk}{N}} & \mbox{for $-N_{\mathrm{CP}}\!\leq\! n\!\leq \!N\!\!-\!\!1\!$}\\
0 & \mbox{otherwise}
\end{array}
\right.
\label{eq_IFFT}
\end{equation}
where $x^{(0)}_{n}$ for $n=0,...,N-1$ denotes preamble samples (RP samples) without the CP. 
Without the loss of generality we narrow our considerations to one OFDM frame. The transmitted signal $\tilde{x}(n)$ composed of subsequent NC-OFDM symbols is thus:
\begin{equation}
\tilde{x}(n)=\sum_{p=0}^{P-1}x^{(p)}_{n-p(N+N_{\mathrm{CP}})}.
\label{eq_frame}
\end{equation}
The signal observed at the NC-OFDM receiver is distorted by the $L$-path fading channel with $l$-th path channel coefficient $h(l)$, the CFO normalized to SCs spacing $\nu$, the additive interference $i(n)$ and the white noise $w(n)$. 
The $n$-th sample of the received signal $r(n)$ equals:
\begin{equation}
r(n)= \sum_{l=0}^{L-1}\tilde{x}(n-l)h(l)
e^{j2\pi \frac{n\nu}{N}}+i(n)+w(n).
\label{eq_received_signal}
\end{equation}
Note that discrete-time representation is sufficient for our considerations. It was shown in \cite{Morelli_07_overview_synchr} that the timing error lower than the sampling period is compensated by an NC-OFDM equalizer. 
\section{Proposed synchronization algorithm: LUISA}
\begin{figure}[tbp]
\centering
\includegraphics[width=3.5in]{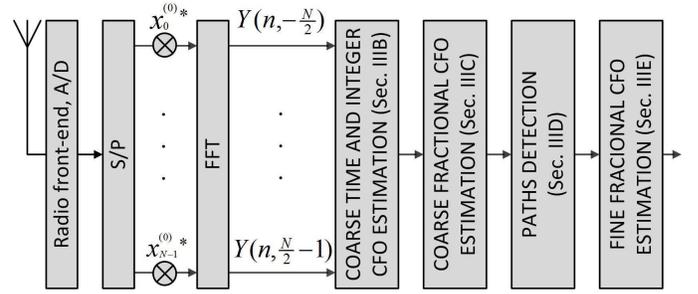}
\caption{NC-OFDM receiver with the proposed synchronization algorithm.}
\label{fig_receiver_diagram}
\end{figure}
The conceptual diagram of LUISA is shown in Fig.\ref{fig_receiver_diagram}. At the output of the radio front-end and the analog-to-digital converter, the baseband complex samples undergo serial-to-parallel (S/P) conversion. The RP is assumed to be known at the receiver like other transmission parameters, e.g. through a suitable \emph{randezvous} technique or a control channel. The RP can be generated at the receiver from a pseudo-random generator, with the same initial state as the one at the transmitter. Note that in the case of perfect time and frequency synchronization, the received preamble samples are correlated with the RP, which results in high amplitude peak of the calculated cross-correlation. Moreover, the SNR at the output of cross-correlator is maximized in the case of AWGN channel \cite{Oberg_2001_modulation_detection_coding} (In Sec. \ref{sec_statistical_properties} it will be shown that multipath propagation degrades the performance slightly by its contribution to noise-power at the correlator output. Moreover, the interference spectrum non-overlapping with the occupied NC-OFDM subcarriers is significantly attenuated by the cross-correlation operation.) 
However, even for correct timing, correlation of the received signal with the RP would not result in any amplitude peak, in the case of relatively high CFO. This is because each cross-correlation component would be added with a different phase, dependent on the CFO. In order to counteract this effect, $N$ consecutive input samples are multiplied by the complex conjugates of the RP and fed to the FFT block.
Thus, at the $k$-th 
output of FFT at the $n$-th sampling moment we obtain: 

\begin{equation}
Y(n,k)=\sum_{m=0}^{N-1}r(n+m)x^{(0)*}_{m}e^{-j2\pi\frac{mk}{N}}\mathrm{,}
\label{eq_Y_k_n}
\end{equation}
where $Y(n,k)$ will be called the \emph{synchronization variable} in the remainder of the paper and $(\cdot)^{*}$ denotes complex conjugate.
It can be calculated in the frequency domain as
\begin{equation}
Y(n,k)=\sum_{k1\in \mathbf{I_{\mathrm{RP}}}}
d_{k1}^{(0)*}R(n,(k+k1)^{\circ})\mathrm{,}
\label{eq_Y_k_n_freq}
\end{equation}
where $R(n,k)$ is an FFT result of $r(n+m)$ for $m=0,...,N-1$ at normalized frequencies $k\in\{-N/2,...,N/2-1\}$, and $(k)^{\circ}$ denotes cyclic indexing of elements in a vector. Observe that (\ref{eq_Y_k_n_freq}) is the frequency-domain equivalent of the cross-correlation of RP and the received signal for different frequency shifts $k$.  
 By substituting (\ref{eq_received_signal}) to (\ref{eq_Y_k_n}) we obtain:
\begin{align}
Y(n,k)=&\sum_{l=0}^{L-1}h(l)Y_{l}(n,k)+Y_{\mathrm{in}}(n,k)+Y_{\mathrm{no}}(n,k)\mathrm{,}
\label{eq_Y_k_n2}
\end{align}
where
\begin{equation}
Y_{l}(n,k)=e^{j2\pi \frac{n\nu}{N}} \sum_{m=0}^{N-1}\tilde{x}(n+m-l)x^{(0)*}_{m}e^{j2\pi\frac{m(\nu-k)}{N}}\mathrm{,}
\label{eq_Y_l_k_n2}
\end{equation}
\begin{equation}
Y_{\mathrm{in}}(n,k)=\sum_{m=0}^{N-1}i(n+m)
x^{(0)*}_{m}e^{-j2\pi \frac{mk}{N}} \mathrm{,}
\label{eq_Y_i_k_n2}
\end{equation}
\begin{equation}
Y_{\mathrm{no}}(n,k)=\sum_{m=0}^{N-1}w(n+m)
x^{(0)*}_{m}e^{-j2\pi \frac{mk}{N}} \mathrm{,}
\label{eq_Y_i_k_n3}
\end{equation}
and the defined $Y_{l}(n,k)$, $Y_{\mathrm{in}}(n,k)$ and $Y_{\mathrm{no}}(n,k)$ are the $l$-th channel-path component, the interference- and the noise involving components of $Y(n,k)$ respectively. The next steps of the proposed synchronization algorithm will be explained in further subsections (\ref{sec_coarse_time}--\ref{sec_fine_frac_CFO}) as indicated in respective blocks in Fig. \ref{fig_receiver_diagram}, after we discuss statistical properties of $Y(n,k)$ in subsection \ref{sec_statistical_properties}.

\subsection{Statistical properties of the synchronization variable}
\label{sec_statistical_properties}

In this paper, we consider two preamble-shapes generated by the pseudo-random $d_{k}^{(0)}$ symbols modulating SCs. The \emph{S\&C preamble}
utilizes only even indices out of set $\mathbf{I}$ (SCs of odd incides are modulated by zeros), i.e., $\mathbf{I_{\mathrm{RP}}}\subset \mathbf{I}$. In order to keep the constant power of the transmitted signal, vector $x_{n}^{(0)}$ for $n=-N_{\mathrm{CP}},...,N-1$ is multiplied by $\sqrt{2}$. 
Moreover, we also consider a preamble generated by modulation of all $\alpha$ SCs by the non-zero pseudo-random symbols, i.e., $\mathbf{I_{\mathrm{RP}}}= \mathbf{I}$. It is referred to as a \emph{simple preamble}. 

According to the Central Limit Theorem (CLT), the samples $\tilde{x}(n)$ can be treated as complex random Gaussian variables with the zero mean and variance  $\sigma_{\mathrm{x}}^{2}$, approximately independent for sufficiently high $\alpha$ (for $1\ll \alpha<N$), as shown in \cite{Wei_2010_rozklad_OFDM} (with the exception for repeatability of the CP or both parts of the S\&C preamble, which will be taken into account in latter calculations). 

The noise $w(n)$ is modeled as Gaussian complex random variable of the zero-mean and variance $\sigma_{\mathrm{w}}^{2}$. 
The in-band interference observed in the NC-OFDM system 
is modeled as:  
\begin{align}
i(n+m)=\frac{1}{\sqrt{N}}\sum_{k=-N/2}^{N/2-1}g_{k}(n)e^{j2\pi \frac{mk}{N}}\mathrm{,}
\label{eq_interf_freq}
\end{align}
where
$g_{k}(n)$ is the discrete-frequency representation of the discrete interference signal at the $k$-th SC
at the $n$th moment, 
and $m\in\{0,...,N-1\}$. Importantly, this representation of (\ref{eq_interf_freq}) holds for any interfering signal, because the $N$-size DFT coefficients $g_{k}(n)$ can perfectly represent any series of $N$ samples of $i(n+m)$, including the interference having the spectrum non-overlapping with the NC-OFDM waveform.
Note that according to the Parseval theorem: $\sum_{k=-N/2}^{N/2-1}\left|g_{k}(n)\right|^{2}=N\sigma_{\mathrm{i}}^{2}$, where $\sigma_{\mathrm{i}}^{2}$ is the interference power.
Moreover, the LU-generated interference has usually most of the energy located in the NC-OFDM spectrum notch(es), i.e. in frequency bands not utilized by NC-OFDM (in equation (\ref{eq_interf_freq}), $g_{k}(n)\approx 0$ for $k\in \mathbf{I}$). As shown in \cite{Yucek07},  $g_{k}(n)$ can be modeled as a random complex Gaussian variable with the zero mean, uncorrelated with $g_{k'}(n)$ for $k\neq k'$, although interference variances for $k$ and $k'$ can be correlated.   

As shown above, the $r(n)$ sample for each $n$ can be treated as an approximately independent complex random variable. Based on the CLT, $Y(n,k)$ probability distribution can be modeled as complex Gaussian, because it is a sum of $N$ weighted random variables.
The expected value of $\mathbb{E}[Y(n,k)]$ is calculated in Appendix \ref{sec_appendix_mean_and_var_Y_n_k}, giving:
\begin{align}
&\mathbb{E}[Y(n,k)]=\sum_{l=0}^{L-1}h(l)\mathbb{E}[Y_{l}(n,k)], 
\label{eq_exp_Y_k_n_minimized}
\end{align}
where
\begin{align}
\mathbb{E}[Y_{l}(n,k)]=e^{j2\pi \frac{n\nu}{N} }\sum_{m=0}^{N-1}\mathbb{E}[\tilde{x}(n+m-l)x^{(0)*}_{m}]e^{j2\pi\frac{m(\nu-k)}{N}}\mathrm{.}
\label{eq_exp_Y_k_n_l}
\end{align}
\begin{figure}[t]
\centering
\includegraphics[width=3.5in]{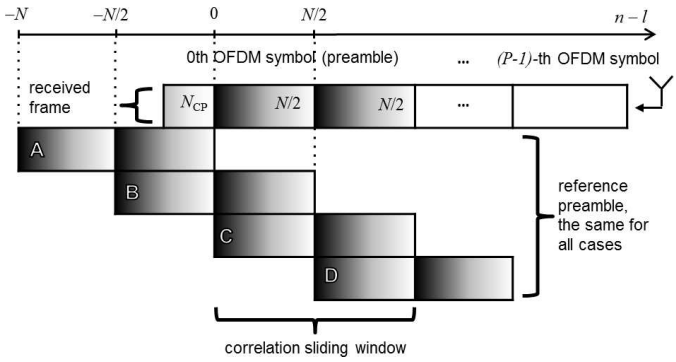}
\caption{Cases of non-zero correlation when preamble consists of two repeated sequences of $N/2$ samples as in the S\&C algorithm. The shade of grey represents the same preamble sample (either originally transmitted 
or in the RP at the receiver). }
\label{fig_corr_cases}
\end{figure}
Note that $\mathbb{E}[\tilde{x}(n+m-l)x^{(0)*}_{m}]$ is non-zero only when $\tilde{x}(n+m-l)$ and $x^{(0)*}_{m}$ are correlated. All cases when this is true are presented in Fig. \ref{fig_corr_cases} for different values of $n-l$. 
For Fig. \ref{fig_corr_cases}, it is assumed that the S\&C preamble is used. 
The received signal (at the top of Fig. \ref{fig_corr_cases}) is processed according to (\ref{eq_Y_k_n}), where the correlation window of $N$ samples for every time instance $n$ is applied. As the correlation sliding window shifts in time over the received signal samples (to the right in Fig. \ref{fig_corr_cases}), there are a few cases when $\tilde{x}(n+m-l)$ and $x^{(0)}_{m}$ are highly correlated: \emph{a}) when only the CP of the preamble is received (Case A); \emph{b}) when the second half of the RP is aligned with the first part of the received preamble inside the correlation sliding window, and additionally $N_{\mathrm{CP}}$ samples of the received preamble CP are aligned with end of the first part of the RP (Case B); \emph{c}) when all $N$ samples of the RP are aligned with the received preamble samples (Case C); \emph{d}) when the second half of the received preamble is aligned with the first half of the RP (Case D).
In case of the simple preamble only cases A and C 
result in $\mathbb{E}[\tilde{x}(n+m-l)x^{(0)*}_{m}]\neq 0$ as the simple preamble does not have two identical halves. The expected values $\mathbb{E} [Y_{l}(n,k)]$ are presented in Table \ref{table_variance_mean} and the calculation example for Case C can be found in Appendix \ref{sec_appendix_mean}.

\begin{figure}[tb]
\centering
\includegraphics[width=3.5in]{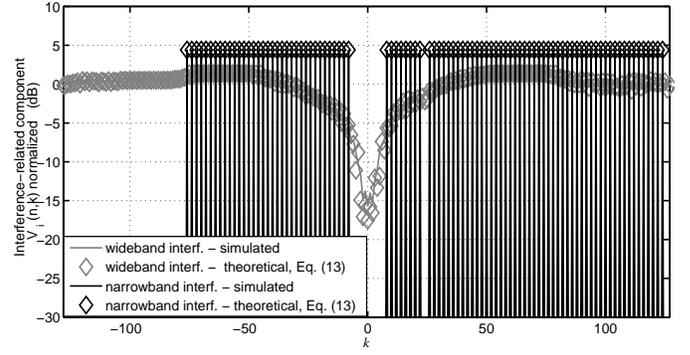}
\caption{ Interference-related component of $\mathbb{V}\left[ Y(n,k)\right]$: $V_{\mathrm{i}}(n,k)$ 
normalized over $N\sigma^{2}_{\mathrm{i}}\sigma^{2}_{\mathrm{x}}$.
}
\label{fig_variance_of_interf}
\end{figure} 
Derivation of the variance of $Y(n,k)$, $\mathbb{V}[Y(n,k)]$, can be found in Appendix \ref{sec_appendix_mean_and_var_Y_n_k}, resulting in:
\begin{align}
\mathbb{V}\left[Y(n,k)\right]=&\sum_{l=0}^{L-1}|h(l)|^{2}\mathbb{V}\left[ Y_{l}(n,k)\right]
\label{eq_variance_accurate_interf_final_kopia}
 \\&\nonumber
+\sum_{k_{1}=0}^{N-1}
\mathbb{E}[ 
|g_{k1}(n)|^{2}]\mathbb{E}[|d^{(0)}_{(k1-k)^{\circ}}|^{2}]
+N\sigma^{2}_{\mathrm{w}}\sigma^{2}_{\mathrm{x}}.
\end{align}
Let us denote the interference-related component in (\ref{eq_variance_accurate_interf_final_kopia}) as: $\sum_{k_{1}=0}^{N-1} \mathbb{E}[ 
|g_{k1}(n)|^{2}]\mathbb{E}[|d^{(0)}_{(k1-k)^{\circ}}|^{2}]=$ $V_{\mathrm{i}}(n,k).$ 
For the interference and NC-OFDM signal equally distributed in the whole band,  $V_{\mathrm{i}}(n,k)=N\sigma^{2}_{\mathrm{i}}\sigma^{2}_{\mathrm{x}}$. 
Typically, the interference occurs in a different band than the NC-OFDM spectrum, and for $k\approx 0$ (small expected CFO), $V_{\mathrm{i}}(n,k)$ is small or in case of both signals being orthogonal - equal to zero. (It proofs that LUISA utilizes the NC-OFDM properties, and is able to reject the LU-originating interference). Example plots of this interference-related variance component being normalized to $N\sigma^{2}_{\mathrm{i}}\sigma^{2}_{\mathrm{x}}$ for the interference models used in the simulations are shown in Fig. \ref{fig_variance_of_interf} 
(observe small values for $k\approx 0$). However,
$V_{\mathrm{i}}(n,k)$ 
can be higher if the useful signal and the interference overlap in frequency. It can reach a maximum of $N^{2}\sigma^{2}_{\mathrm{i}}\sigma^{2}_{\mathrm{x}}$ for the theoretical case of both interference and NC-OFDM signal being complex sinusoids. Thus, $N\sigma^{2}_{\mathrm{i}}\sigma^{2}_{\mathrm{x}}$ reflects the intermediate worst case.     

The variances $\mathbb{V}[Y_{l}(n,k)]$ are presented in Table \ref{table_variance_mean}, and derived for Case C in Appendix \ref{sec_appendix_variance}.
\setlength{\tabcolsep}{2pt}
\renewcommand{\arraystretch}{1.1}
\linespread{1}
\begin{table*}[htb]
\caption{Means and variances of the channel path components  $Y_{l}(n,k)$ of the synchronization variable $Y(n,k)$}
\label{table_variance_mean}
\centering
\begin{tabular}{|@{}c@{}|c|@{}c@{}|@{}c@{}|}
\hline
\textbf{Case} & \textbf{preamble}   & $\mathbb{E} [Y_{l}(n,k)]$ & $\mathbb{V}[Y_{l}(n,k)]$
 \\ \hline
\multirow{2}{*}{\begin{tabular}[c]{@{}c@{}}A\\ $n-l=-N$\\ $m=\{N-N_{\mathrm{CP}},...,N-1\}$\end{tabular}}
 & simple & \multirow{2}{*}[0ex]{
\begin{tabular}[c]{@{}c@{}}$\sigma^{2}_{\mathrm{x}}e^{j2\pi (l-N)\frac{\nu}{N}}e^{j\pi \left(\nu-k\right)\left(2-\frac{N_{\mathrm{CP}}+1}{N}\right)}\cdot$\\ $\frac{\sin\left(\pi N_{\mathrm{CP}}\frac{\nu -k}{N} \right)}{\sin\left(\pi \frac{\nu -k}{N} \right)}$\end{tabular}}
& \multirow{2}{*}[-2ex]{$\sigma^{4}_{\mathrm{x}} N_{\mathrm{CP}}$} \\[3ex] \cline{2-2}
& S\&C   &  &  \\[2ex] \hline
\begin{tabular}[c]{@{}c@{}}B\\ $n-l=-\frac{N}{2}$\\ $m=\{\frac{N}{2}-N_{\mathrm{CP}},...,N-1\}$\end{tabular}
 & S\&C   
& 
\begin{tabular}[c]{@{}c@{}}$\sigma^{2}_{\mathrm{x}}e^{j2\pi (l-\frac{N}{2})\frac{\nu}{N}}e^{j\pi \left(\nu-k\right)\left(\frac{3}{2}-\frac{N_{CP}+1}{N}\right)}\cdot$\\ 
$\frac{\sin\left(\pi \left( \frac{1}{2}+\frac{N_{\mathrm{CP}}}{N}\right) (\nu -k)\right)}{\sin\left(\pi \frac{\nu -k}{N} \right)}$
\end{tabular}
&
$\sigma^{4}_{\mathrm{x}}\frac{N}{2}+\left(1+2\cos(\pi(\nu-k))\right)\sigma^{4}_{\mathrm{x}}N_{\mathrm{CP}}$ \\[4ex] \hline
\multirow{2}{*}{\begin{tabular}[c]{@{}c@{}}C\\ $n-l=0$\\ $m=\{0,...,N-1\}$\end{tabular}} & simple
& \multirow{2}{*}[-2ex]{$\sigma^{2}_{\mathrm{x}}e^{j2\pi l\frac{\nu}{N}}e^{j\pi \left(\nu-k\right)\left(1-\frac{1}{N}\right)}\frac{\sin\left(\pi (\nu-k)  \right)}{\sin\left(\pi \frac{\nu -k}{N} \right)}$} 
& $\sigma^{4}_{\mathrm{x}} N$ \\[3ex] \cline{2-2} \cline{4-4} 
 & S\&C &   & $\left(1+\cos(\pi(\nu-k))\right)N\sigma^{4}_{\mathrm{x}}$ \\[2ex] \hline
\begin{tabular}[c]{@{}c@{}}D\\ $n-l=\frac{N}{2}$\\ $m=\{0,...,\frac{N}{2}-1\}$\end{tabular}
& S\&C & $\sigma^{2}_{\mathrm{x}}e^{j2\pi (l-\frac{N}{2})\frac{\nu}{N}}e^{j\pi \left(\nu-k\right)\left(\frac{1}{2}-\frac{1}{N}\right)}\frac{\sin\left(\frac{\pi}{2} (\nu-k)  \right)}{\sin\left(\pi \frac{\nu -k}{N} \right)}$ & $\sigma^{4}_{\mathrm{x}}N$ \\[4ex] \hline
\multicolumn{2}{|@{}c@{}|}{otherwise} & 0 & $\sigma^{4}_{\mathrm{x}}\min\left(\max \left(0,N+N_{\mathrm{CP}}+n-l \right) ,N\right)$   \\ \hline
\end{tabular}
\end{table*}
\linespread{1.55}
\subsection{Coarse time-offset and integer CFO estimation}
\label{sec_coarse_time}
\begin{figure}[!t]
\centering
\includegraphics[width=3.5in]{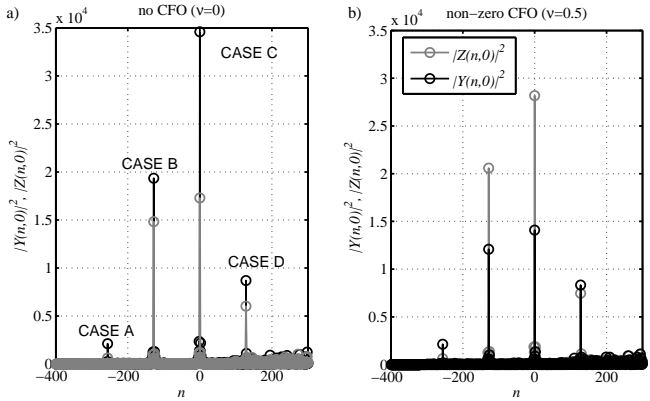}
\caption{$|Y(n,k)|^{2}$ and $|Z(n,k)|^{2}$ for $k=0$, S\&C preamble, $N=256$, $N_{\mathrm{CP}}=N/4$, no interference, no noise, no multipath ($L=1$), and no CFO ($\nu=0$) (a) and worst case of CFO ($\nu=0.5$) (b) considered.}
\label{fig_EXAMPLE_Y}
\end{figure}

On the basis of the defined synchronization variable $Y(n,k)$, the coarse time and frequency synchronization can be directly obtained. The search for the peak $Y(n,k)$ power must be made in two dimensions: over frequency and time. The simplest decision method would seek $n=n_{\mathrm{Y}}$ and $k=k_{\mathrm{Y}}$ time- and frequency indices to maximize the squared absolute of $Y(n,k)$:
\footnote{Although, the first stage of synchronization is obtained by maximization (as in most competitive algorithms), LUISA requires the threshold 
in order to detect the absence of preamble in the observed range of samples $n$. 
It can be obtained as 
in Section \ref{sec_paths_detection} with the probability of false detection
divided by the total number of points in the 2D plane (instead of $2N_{\mathrm{CP}}$). Moreover, at least $N$ samples following the sample exceeding this threshold should be examined to find the highest peak (Case C).} 
\begin{equation}
[n_{\mathrm{Y}},k_{\mathrm{Y}}]=\argmax_{n,k} |Y(n,k)|^{2}.
\label{eq_simple_detection}
\end{equation}
An example plot of $|Y(n,k)|^{2}$ versus $n$ for $k=0$ in the idealistic condition of no-noise, no-interference and no-multipath effect can be observed in Fig. \ref{fig_EXAMPLE_Y} for $N=256$, S\&C preamble and $N_{\mathrm{CP}}=N/4$. There are four peaks of $|Y(n,k)|^{2}$, which reflect four cases when non-zero expected value of $Y(n,k)$ is obtained, as presented in Fig. \ref{fig_corr_cases}. 
According to the statistical properties of the complex random Gaussian variable $Y_{l}(n,k)$ given in Table \ref{table_variance_mean}, it is the most probable that the strongest path in case C, i.e. $n_{\mathrm{Y}}=l_{\max}=\argmax_{l}|h_{l}|^{2}$, will be found with (\ref{eq_simple_detection}) because for $n_{\mathrm{Y}}$ the magnitude of the expected value of $Y(n_{Y},k_{Y})$ should be the highest. 
This expected value for $\sigma^{2}_{\mathrm{w}}=0$, $\sigma^{2}_{\mathrm{i}}=0$ and $n_{\mathrm{Y}}=l_{\max}$ equals:
\begin{equation}
\mathbb{E}[|Y(n_{\mathrm{Y}},k_{\mathrm{Y}})|^{2}]\approx|h_{l_{\max}}|^{2}\sigma^{4}_{\mathrm{x}}\left(\!\frac{\sin(\pi (\nu-k_{Y}))}{\sin(\pi (\nu -k_{Y})/N)}\!\right)^{2}.
\label{eq_simple_detection_expectation}
\end{equation}
Note that  $\mathbb{E}[|Y(n_{\mathrm{Y}},k_{\mathrm{Y}})|^{2}]=\mathbb{V}[Y(n_{\mathrm{Y}},k_{\mathrm{Y}})]+|\mathbb{E}[Y(n_{\mathrm{Y}},k_{\mathrm{Y}})]|^{2}$, but because $|\mathbb{E}[Y(n_{\mathrm{Y}},k_{\mathrm{Y}})]|^{2}$ is up to $N$ times higher than $\mathbb{V}[Y(n_{\mathrm{Y}},k_{\mathrm{Y}})]$ for high SNR and for non-overlapping spectra of interference and the NC-OFDM signal (according to (\ref{eq_variance_accurate_interf_final_kopia}) and Table \ref{table_variance_mean}), it can be approximated by: $\mathbb{E}[|Y(n_{\mathrm{Y}},k_{\mathrm{Y}})|^{2}]\approx|\mathbb{E}[Y(n_{\mathrm{Y}},k_{\mathrm{Y}})]|^{2}$.  
 
\subsubsection{A method to increase the CFO robustness}

The search for optimum $k_{\mathrm{Y}}$ in (\ref{eq_simple_detection}) for $n_{\mathrm{Y}}=l_{\max}$ is done over $k=-N/2,...,N/2-1$. Let us note that there exists one FFT output, for which the error of the CFO estimate is equal or lower than half of the SCs spacing, i.e. $|\nu-k|\leq 0.5$. This FFT-output index should be chosen by (\ref{eq_simple_detection}) as $k_{\mathrm{Y}}$ because it maximizes the expected value in (\ref{eq_simple_detection_expectation}). Assuming $|\nu-k_{\mathrm{Y}}|\ll N$, the denominator in (\ref{eq_simple_detection_expectation}) can be approximated by $\sin(\pi (\nu -k_{\mathrm{Y}})/N)\approx\pi (\nu -k_{\mathrm{Y}})/N$, and $\mathbb{E}[|Y(n_{\mathrm{Y}},k_{\mathrm{Y}})|^{2}]$ reaches its maximum equal to $|h_{l_{max}}|^{2}\sigma^{4}_{\mathrm{x}}N^{2}$ for $|\nu-k_{\mathrm{Y}}|=0$ (example shown in Fig. \ref{fig_EXAMPLE_Y}a). The minimum over the whole domain, i.e. $\nu-k_{\mathrm{Y}}\in\left<-0.5;0.5\right>$, is obtained for $|\nu-k_{\mathrm{Y}}|=0.5$, and equals $|h_{l_{max}}|^{2}\sigma^{4}_{\mathrm{x}}\left(N2/\pi\right)^{2}$, which is around $40\%$ of the maximum (example shown in Fig. \ref{fig_EXAMPLE_Y} b). This may cause a decrease of the correct time-frequency-point detection probability for the CFO not aligned with the frequencies of the FFT used to calculate $Y(n,k)$. To increase the detection probability in this case, we can utilize two adjacent FFT outputs: $Y(n,k)$ and $Y(n,k+1)$. For $\nu-k= 0.5$, the synchronization variable $Y(l,k+1)$ 
has the same absolute value of the mean as $Y(l,k)$: $|\mathbb{E}[Y(l,k+1)]|=|\mathbb{E}[Y(l,k)]|$. Both components can be added in phase if $Y(l,k+1)$ is corrected by factor: $-exp(-j\pi/N)$ (according to Table \ref{table_variance_mean}) maximizing the expected value for optimal timing point and $\nu-k= 0.5$. 
Thus, let $Z(n,k)=\left[ Y(n,k)-Y(n,k+1)exp(-j\pi/N)\right] /\sqrt{2}$ (multiplier $\frac{1}{\sqrt{2}}$ keeps the same variance of $Z(n,k)$ as of $Y(n,k)$, and the same probability of false detection), and:
\begin{equation}
[n_{\mathrm{Z}},k_{\mathrm{Z}}]=\arg\max_{n,k} |Z(n,k)|^{2}.
\label{eq_max_Z}
\end{equation}
The calculation of $Z(n,k)$ is based on the $Y(n,k)$ values, resulting in a small increase of computational complexity in comparison to detection based on (\ref{eq_simple_detection}) only. Following the assumption that $l_{max}=n$, that $Y(n,k)$ and $Y(n,k+1)$ are uncorrelated \cite{Candan_freq_est2013}, and that $\mathbb{E}[|Z(n,k)|^{2}]\approx |\mathbb{E}[Z(n,k)]|^{2}$ 
(as in (\ref{eq_simple_detection_expectation})) the mean value of $|Z(n,k)|^{2}$ can be obtained using Table \ref{table_variance_mean}:
\begin{align}
&\mathbb{E}\left[|Z(n,k)|^{2}\right]=
\frac{1}{2}\bigl|\mathbb{E}\left[Y(n,k)\right]-\mathbb{E}\left[Y(n,k+1)
\right]e^{-j\frac{\pi}{N}}\bigr|^{2}=
\nonumber
\\&
\frac{1}{2}|h_{l_{max}}|^{2}\sigma^{4}_{\mathrm{x}}\!\!\left(\!\!\frac{\sin(\pi (\nu-k))}{\sin(\pi (\nu\! -\!k)/N)}\!-\!\frac{\sin(\pi (\nu-k))}{\sin(\pi (\nu\!\! -\!\!k\!\!-\!\!1)/N)}\!\right)^{2}\!\!\!.
\label{eq_approx_Y}
\end{align}  
Similarly, for both the simple- and the S\&C preamble cases, the variance of $Z(n,k)$ can be calculated and it equals the variance of $Y(n,k)$ in the case of a simple preamble.      
Both summation components under square in (\ref{eq_approx_Y}) have the same phase for $\nu-k\approx0.5$ which increases detection probability. For $\nu-k=0.5$, the decision variable in (\ref{eq_approx_Y}) can be approximated as $0.5|h_{l_{max}}|^{2}\sigma^{4}_{\mathrm{x}}\left(N4/\pi\right)^{2}$, which is twice the minimum (obtained for $\nu-k=0.5$) and about $80\%$ of the maximum (obtained for $\nu-k=0$) of $\mathbb{E}[|Y(n_{\mathrm{Y}},k_{\mathrm{Y}})|^{2}]$. (The example of this decision metric is shown in Fig. \ref{fig_EXAMPLE_Y}.) Thus, 
the maximum out of both decision variables maxima is chosen:
\begin{equation}
[n_{\mathrm{M}},k_{\mathrm{M}}]=\!\!\!\argmax_{(n_{\mathrm{Y}},k_{\mathrm{Y}}), (n_{\mathrm{Z}},k_{\mathrm{Z}})}\!\!\!\! \left(|Y(n_{\mathrm{Y}},k_{\mathrm{Y}})|^{2},|Z(n_{\mathrm{Z}},k_{\mathrm{Z}})|^{2}\right). 
\label{eq_improved_detection}
\end{equation}
The variable $|Y(n_{\mathrm{Y}},k_{\mathrm{Y}})|^{2}$ is higher than $|Z(n_{\mathrm{Z}},k_{\mathrm{Z}})|^{2}$ for the CFOs closer to the frequencies of the receiver FFT, otherwise $|Z(n_{\mathrm{Z}},k_{\mathrm{Z}})|^{2}$ should be higher.
Note that if (\ref{eq_max_Z}) finds the coarse time and frequency synchronization point, i.e. $|Z(n_{\mathrm{Z}},k_{\mathrm{Z}})|^{2} >|Y(n_{\mathrm{Y}},k_{\mathrm{Y}})|^{2}$, it is ambiguous whether a peak occurs for $\nu$ closer to the normalized frequency $k_{\mathrm{Z}}$ or to $k_{\mathrm{Z}}+1$ ($\nu\in (k_{\mathrm{Z}};k_{\mathrm{Z}}+1)$). 
In this case, and for $|Y(n_{\mathrm{Z}},k_{\mathrm{Z}}+1)|^{2}>|Y(n_{\mathrm{Z}},k_{\mathrm{Z}})|^{2}$, $k_{\mathrm{M}}$ is modified to $k_{\mathrm{Z}}+1$.

In the reminder of this paper, we use $[n_{\mathrm{M}},k_{\mathrm{M}}]$ as a result of the coarse time and integer CFO estimation.
Note that the computational complexity of this estimation can be decreased if the receiver has the knowledge on the upper limit of the CFO $\nu_{\mathrm{M}}$, where  $|\nu|<\nu_{\mathrm{M}}$. It allows for limiting the integer-CFO search range to $k=\{-\lceil\nu_{\mathrm{M}}\rceil,...,\lceil\nu_{\mathrm{M}}\rceil\}$, where $\lceil\cdot\rceil$ is the \emph{ceil} function.

\subsection{Coarse fractional CFO estimation}
As shown above,
$Y(n,k)$ is a Gaussian random variable with non-zero mean when preamble is detected. The mean value has $sinc$-like shape (over $k$). Thus, for a given $n$, $Y(n,k)$ is similar to the FFT output if the input signal was the complex sinusoid of an unknown frequency with the AWGN \cite{Candan_freq_est2013}. The algorithm to find this frequency presented in \cite{Candan_freq_est2013} can be used to estimate $\nu$ corresponding to the peak of $sinc$-like frequency response. 
The integer CFO estimate $k_{\mathrm{M}}$ is used to calculate the coarse CFO estimate $\widehat{\nu_{\mathrm{C}}}$ using the following formula:
\begin{equation}
\widehat{\nu_{\mathrm{C}}}=\frac{N}{\pi} \mathrm{atan} \left( 
\tan \left(\frac{\pi}{N}\right)
\Re\left(\frac{Q_{1}(n_{\mathrm{M}},k_{\mathrm{M}})}{Q_{2}(n_{\mathrm{M}},k_{\mathrm{M}})}
\right)
\right),
\label{eq_freq_est_first}
\end{equation} 
where $\Re(\cdot )$ is the real part of a complex-number argument,
$Q_{1}(n,k)=Y(n,k-1)-Y(n,k+1)$,
$Q_{2}(n,k)=2Y(n,k)-Y(n,k-1)-Y(n,k+1)$,
and $\mathrm{tan}(\cdot)$ and $\mathrm{atan}(\cdot)$ are tangent and arctangent functions, respectively. The first (coarse) estimate of $\nu$ is $\widehat{\nu}_{0}=\widehat{\nu}_{\mathrm{C}}+k_{\mathrm{M}}$, i.e., $\widehat{\nu}_{0}\approx \nu$.
The lower index $0$ of $\widehat{\nu}_{0}$ denotes $0-th$ iteration of the CFO estimate, i.e. initial estimate.

\subsection{Paths detection}
\label{sec_paths_detection}
If in the previous steps, the beginning of the maximum-power channel path ($n_{\mathrm{M}}=l_{\max}$) has been found, and the CFO estimate is close to its actual value $\nu$, $Y(n,k)$ can be calculated for $k=\widehat{\nu}_{0}$ using (\ref{eq_Y_k_n}). It increases $|\mathbb{E}[Y_{l}(l,k)]|$ as $\nu-k\approx 0$ (see Table \ref{table_variance_mean}), so that multipath components  can be better distinguished from the noise and interference. In OFDM, all path-components are typically spread over the maximum of $N_{\mathrm{CP}}$ sampling intervals, so $n\in \{n_{\mathrm{M}}-N_{\mathrm{CP}},...,n_{\mathrm{M}}-1,n_{\mathrm{M}}+1,...,n_{\mathrm{M}}+N_{\mathrm{CP}}\}$.  The aim of this step is to find the first and other paths to be used for fine CFO correction. After coarse CFO estimation, the set of detected-paths indices $\mathbf{D}$ consists of the index of the strongest path: $\mathbf{D}=\{n_{\mathrm{M}}\}$. At the path-detection stage, the $n$th received sample represents a path component and is added to set $\mathbf{D}$ as in \cite{Awoseyila_synchronization} if 
\begin{equation}
\left|Y(n,\widehat{\nu}_{0})\right|^{2}>-\sigma_{\mathrm{thr}}^{2}(n)\ln(P_{\mathrm{FD}}/(2N_{\mathrm{CP}})),
\label{eq_threshold}
\end{equation}   
where $P_{\mathrm{FD}}$ is the probability of false detection over $2N_{\mathrm{CP}}$ samples, and 
$\sigma_{\mathrm{thr}}^{2}(n)$ is the estimated input-signal variance for RP not correlated with the received samples (not case A-D). The false detection occurs if $\left|Y(n,\widehat{\nu}_{0})\right|^{2}$ exceeds the threshold despite $n$ does not represent a path-reception moment (noise or interference is detected as a channel path). 
Assuming that the sampling time $n$ is not aligned with any path-reception moment, $\mathbb{E}[Y(n,\widehat{\nu}_{0})]=0$, and $2\left|Y(n,\widehat{\nu}_{0})\right|^{2}/\sigma_{\mathrm{thr}}^{2}(n)$ has chi-square distribution with $2$ degrees of freedom. 
The threshold in (\ref{eq_threshold}) uses chi-square cumulative distribution function. 
Moreover, $\sigma_{\mathrm{thr}}^{2}(n)$ 
can be estimated recursively: 
\begin{align}
\sigma^{2}_{\mathrm{thr}}(n)=&\sigma^{2}_{\mathrm{x}}\sum_{m=0}^{N-1}|r(n+m)|^{2}
\label{eq_sigma_threshold}
 \\\nonumber
=&\sigma^{2}_{\mathrm{thr}}(n-1)+\sigma^{2}_{\mathrm{x}}(|r(n+N-1)|^{2}-|r(n-1)|^{2}).
\end{align}     
The mean of (\ref{eq_sigma_threshold}) can be calculated by substituting expansion of $r(n+m)$ from (\ref{eq_received_signal}) resulting in:

\begin{align}
&\mathbb{E}[\sigma^{2}_{\mathrm{thr}}(n)]=\sigma^{2}_{\mathrm{x}}\sum_{m=0}^{N-1}\bigg(\mathbb{E}[|i(n+m)|^{2}] 
+\mathbb{E}[|w(n+m)|^{2}]+ 
\nonumber \\&
\sum_{l=0}^{L-1}|h(l)|^{2}\mathbb{E}[|\tilde{x}(n+m-l)|^{2}]\bigg)
 =\sigma^{2}_{\mathrm{x}}( \sigma^{2}_{\mathrm{i}}+\sigma^{2}_{\mathrm{w}})N 
\nonumber \\&
+\sum_{l=0}^{L-1}|h(l)|^{2}\sigma_{\mathrm{x}}^{4}\min (\max (0,N+N_{\mathrm{CP}}+n-l),N)
\label{eq_sigma_estimate}
\end{align} 
which equals $\mathbb{V}[Y(n,k)]$ when interference is equally distributed in the whole receiver band (see Appendix \ref{sec_appendix_mean_and_var_Y_n_k} and Sec. \ref{sec_statistical_properties}), and no channel-path is found (the \textquotedblleft\emph{otherwise}\textquotedblright ~row in Table \ref{table_variance_mean}). 


The above search is made over $2N_{\mathrm{CP}}+1$ elements, while the maximum delay spread can be of $N_{\mathrm{CP}}$ samples.
It is proposed that the estimated channel paths indices from set $\mathbf{D}$ are sorted in the descending channel-path-power order, 
and only the strongest path-components are chosen, which are detected over  $N_{\mathrm{CP}}$ samples.  
At this stage, the final estimate of the first channel-path timing $\dot{n}_{\mathrm{M}}$ can be made. It will be the earliest-arriving channel path component, i.e.,  
\begin{equation}
\dot{n}_{\mathrm{M}}=\min\mathbf{D}.
\end{equation} 
Observe that initial channel estimation can be obtained at this stage as for each $l\in \mathbf{D}$ $\mathbb{E}\left[Y(l,\widehat{\nu}_{0})\right]\approx h(l)N\sigma_{x}^2$. This estimate can be improved later using advanced decision-directed methods \cite{Ozdemir_2007_OFDM_channel_est}.

\subsection{Fine fractional CFO estimation}
\label{sec_fine_frac_CFO}
All detected signal paths-components whose indices are in $\mathbf{D}$ can be used for fine CFO estimation. According to Table \ref{table_variance_mean}, in case C, the expected values of $Y(n,\widehat{\nu}_{0})$ for various $n\in \mathbf{D}$ differ by $h(n)e^{j2\pi \frac{\nu}{N}n}$ while are distorted equally by the CFO. Thus, the method from 
\cite{Candan_freq_est2013} can be modified to use all path-components coherently, and to update $\widehat{\nu}_{0}$ as follows: 
\begin{align}
&\!\Delta \widehat{\nu}_{0}\!=\!
\frac{N}{\pi} \mathrm{atan}\! \left(\!
\tan \!\left(\!\frac{\pi}{N}\!\right)\!
\Re\!\left(\!\frac{\sum_{n\in \mathbf{D}} \{Q_{1}(n,\widehat{\nu}_{0})Y^{*}(n,\widehat{\nu}_{0})\}}{\sum_{n\in \mathbf{D}} \{Q_{2}(n,\widehat{\nu}_{0})Y^{*}(n,\widehat{\nu}_{0})\}}\!
\right)\!\right).
\label{eq_fine_freq_est}
\end{align}        
The correctness of formula (\ref{eq_fine_freq_est}) can be justified by calculating the first order approximation of the expected value of (\ref{eq_fine_freq_est}) substituting the expressions from Table \ref{table_variance_mean} for case C, i.e.,

\begin{align}
&\mathbb{E}\left[\Delta \widehat{\nu}_{0}\right]\approx
\frac{N}{\pi} \mathrm{atan} \left(
\tan \left(\frac{\pi}{N}\right)
\Re\Bigl(\frac{\sum_{n\in \mathbf{D}}\{\mathbb{E}[Y^{*}(n,\widehat{\nu}_{0})]}{\sum_{n\in \mathbf{D}}\{\mathbb{E}[Y^{*}(n,\widehat{\nu}_{0})]} \Bigr. \right.\nonumber\\
&\left. \left. \frac{ 
(\mathbb{E}[Y(n,\widehat{\nu}_{0}-1)]-\mathbb{E}[Y(n,\widehat{\nu}_{0}+1)])\}
}{ 
(2\mathbb{E}[Y(n,\widehat{\nu}_{0})]-\mathbb{E}[Y(n,\widehat{\nu}_{0}-1)]-\mathbb{E}[Y(n,\widehat{\nu}_{0}+1)])\}
}
\right) \right)=
\nonumber \\&
= \frac{N}{\pi} \mathrm{atan} \left(
\tan \left(\frac{\pi}{N}\right)
\Re \left(\frac{
\sum_{n\in \mathbf{D}} |h(n)|^2
}{
\sum_{n\in \mathbf{D}} |h(n)|^2
}\cot\left(\frac{\pi}{N} \right)\right. \right. 
\nonumber\\&
\left. \left.\tan\left(\frac{\pi(\nu-\widehat{\nu}_{0})}{N} \right)
\right) \right)=\nu-\widehat{\nu}_{0},
\label{eq_fine_freq_est_expect}
\end{align}    
where $\mathrm{cot}(\cdot)$ is the cotangent function. The updated version of the frequency estimate is: 
\begin{equation}
\widehat{\nu}_{1}=\widehat{\nu}_{0}+\Delta\widehat{\nu}_{0}.
\label{eq_fine_freq_est_update}
\end{equation} 
This estimate can be further improved by iterative calculation of (\ref{eq_fine_freq_est}) and (\ref{eq_fine_freq_est_update}). In the $\gamma$-th iteration, $\widehat{\nu}_{\gamma}$ is calculated based on $\widehat{\nu}_{\gamma-1}$ and $\Delta\widehat{\nu}_{\gamma-1}$. As the estimated CFO $\widehat{\nu}_{\gamma}$ gets closer to $\nu$, the theoretical estimator MSE decreases \cite{Candan_freq_est2013}.
The MSE of frequency estimation is limited in our case by the non-zero variance of the useful NC-OFDM signal (equal zero for complex sinusoid in \cite{Candan_freq_est2013}). However, in case of S\&C preamble, our proposed method of calculating the CFO update, according to (\ref{eq_fine_freq_est}) utilizes the fact that the variance of $Y_l(n,k)$ at the time sample aligned with the channel-path component $l$ and with the frequency distanced from the CFO by one SCs spacing equals zero ($\mathbb{V}[Y_l(l,\nu+1)]=\mathbb{V}[Y_l(l,\nu-1)]=0$). As $\gamma$ increases, $\widehat{\nu}_{\gamma}$ approaches $\nu$, the variance of $Q_{1}(n,\widehat{\nu}_{\gamma-1})$ decreases, increasing the CFO estimation quality, specially at high SNR and SIR. In this case, the variance of this estimator in the AWGN channel is:
$\mathbb{V}[\nu-\widehat{\nu_{\gamma}}]=\frac{1}{4N~\mathrm{SNR}}\mathrm{,}$
the same as in \cite{Candan_freq_est2013}. It is also $1.6$ times higher than Cramer-Rao lower bound of $\frac{6}{4\pi^{2}N~\mathrm{SNR}}$ \cite{Candan_freq_est2013}.   

\section{Performance and complexity evaluation}
\begin{figure}[tbp]
\centering
\includegraphics[width=3.4in]{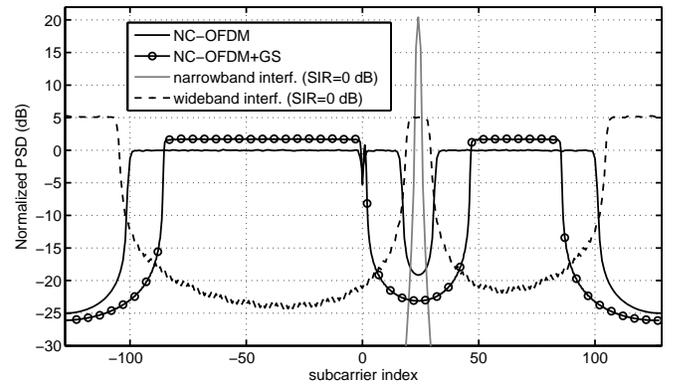}
\caption{Power spectral density of the NC-OFDM signal (with and without guard subcarriers), the in-band NBI at center frequency 24 and in-band OFDM-like WBI occupying bands unused by CR NC-OFDM signal; SIR~=~0~dB, no CFO.}
\label{fig_widmo}
\end{figure}

Below, the results of LUISA are compared against the results of the S\&C algorithm \cite{Schmidl_Cox_1997}, the \emph{Ad-Hoc Detector no. 1} (AHD1) \cite{Sanguinetti_2010_synchr}, and the Ziabari method \cite{Ziabari_synchr_NC_OFDM}.
 In the considered NC-OFDM system, $N=256$ -order IFFT/FFT and the CP of $N_{\mathrm{CP}}=N/16$ samples are applied. 
 We consider three system-scenarios: 1) 
without the application of GSs, 
2) with GSs, and 3) the Dynamic Spectrum Allocation (DSA) with GSs, when the gap in the NC-OFDM spectrum required for the narrowband LU-transmission protection has a random position in frequency for each transmitted frame. 
The interference is simulated in two ways: (i) as the complex sinusoid (as in \cite{Coulson_2004_synchr_narrow_sinusoid}) observed in the NC-OFDM spectrum notch at 24 normalized frequency (strictly NBI) or (ii) as an OFDM-like signal modulated by random QPSK symbols occupying the  band not used by our NC-OFDM signal spectrum, 
and using SCs not orthogonal to our NC-OFDM SCs (wide-band interference -- WBI). 
The example PSD plots of the NC-OFDM signal and the interference having equal power (SIR~=~0~dB) are depicted in Fig. \ref{fig_widmo}.\footnote{Although in this example, SIR=0 dB, the ratio of some SCs power to the interference power observed at their frequencies can be much higher, e.g. up to nearly 25~dB for the WBI.}   
The SNR or SIR are defined as the power ratios 
over the whole NC-OFDM- receiver band.

While both Ziabari and the S\&C method use the S\&C preamble, AHD1 uses the preamble consisting of four identical sequences. The third sequence has the inverted sign as in \cite{Sanguinetti_2010_synchr} which increases the energy leakage into the LU band. LUISA is tested both with the S\&C- and the simple preamble. The preamble and the data symbols use QPSK mapping with random symbols for each simulation run. 
For each case, $10^{5}$ independent OFDM frames have been tested, each consisting of $11$ NC-OFDM symbols, preceded by 2 empty OFDM symbols (interframe period). 
For each OFDM frame, a random instance of 9-path Rayleigh-fading channel is generated according to Extended Vehicular A model \cite{3gpp.36.101}, as well as a random CFO uniformly distributed over $(-3;3)$ normalized to SCs spacing of 15~kHz (as in LTE). As AHD1 has limited CFO estimation range, the simulated range has been limited in this case to $\nu\in (-1.7;1.7)$.
For the S\&C algorithm, the beginning of the preamble is found as a point in the middle between two points achieving $90\%$ of the maximum of timing metric, as proposed in \cite{Schmidl_Cox_1997}, while integer CFO is searched over the range of $\{-20,...,20\}$. The Ziabari method uses its own type of correlation consisting of $4N$ values. It scans the whole set of possible integer CFO values ($N$), while its threshold is set for $P_{\mathrm{FD}}=10^{-5}$ (the same as in LUISA). 
Duration of the window used for the first channel path detection is $5$ samples. 
In LUISA, 
the number of iterations of fine CFO estimation using 
(\ref{eq_fine_freq_est_update}) is set to 2. Finally, AHD1 uses its parameter $\lambda=0.1$ as in \cite{Sanguinetti_2010_synchr}.

\subsection{Simulation results}
\subsubsection{ A system not applying the GSs}
\begin{figure}[tbp]
\centering
\includegraphics[width=3.5in]{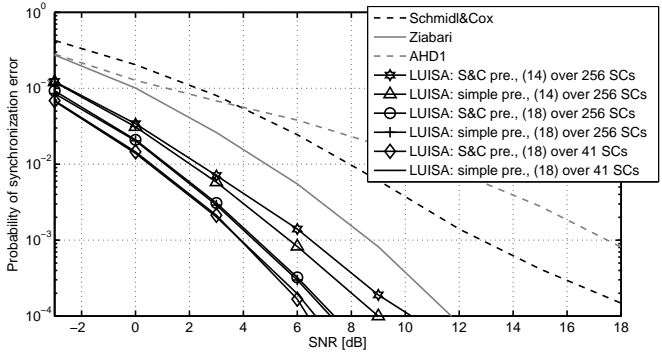}
\caption{Estimated probability of frame synchronization error for LUISA configurations (preamble type, coarse time-frequency point detection method (Equation no.), the range of integer CFO search (no. of SCs); no GSs, no interference.}
\label{fig_prawd_no_interf}
\end{figure}

In this scenario, $\mathbf{I}=\{-100,...,-1,1,...,16,32,...,100\}$. First, the presence of AWGN, CFO and channel fading is considered and no interference. 
The probability of frame synchronization error estimated in the simulations is depicted in Fig. \ref{fig_prawd_no_interf}. The frame is assumed to be correctly synchronized if the error in time is lower than $N_{\mathrm{CP}}$ samples ($\dot{n}_{\mathrm{M}}\in\{-N_{\mathrm{CP}}+1,...,N_{\mathrm{CP}}-1\}$), and the error in frequency is smaller than half of the SCs spacing ($|\widehat{\nu_{2}}-\nu|<0.5$). LUISA in all configurations outperforms the other reference algorithms. For the probability of the synchronization error equal to $10^{-3}$, the S\&C and Ziabari algorithms require SNR of 12.8~dB and 8.7~dB respectively. AHD1 has been designed for interference limited environment, so the requirement of 18~dB of SNR is expected and coherent with the results presented in \cite{Sanguinetti_2010_synchr}. LUISA obtains the same reliability for SNR of only 6.5~dB and 5.7~dB in case of simple detection (using (\ref{eq_simple_detection})) for the S\&C preamble and for the simple preamble respectively, for the full integer-CFO range. 
At higher SNRs $\mathbb{V}[Y(n,k)]$ consists mostly of $\mathbb{V}[Y_{l}(n,k)]$ components  (according to (\ref{eq_variance_accurate_interf_final})), that is higher for the S\&C preamble (as shown in Table \ref{table_variance_mean}) causing higher probability of synchronization error than in the case of simple preamble. 
Additional improvement can be obtained by detecting the frame-beginning time with formula (\ref{eq_improved_detection}) which is more robust against non-integer CFO, and limiting the range of possible CFO to 41 SCs, as in the S\&C algorithm, i.e. $\nu_{\mathrm{M}}=20$. This results in an SNR gain of 8.9~dB, 4.8~dB and 14 dB of LUISA over the S\&C, Ziabari and AHD1 method respectively.
\begin{figure}[tbp]
\centering
\includegraphics[width=3.5in]{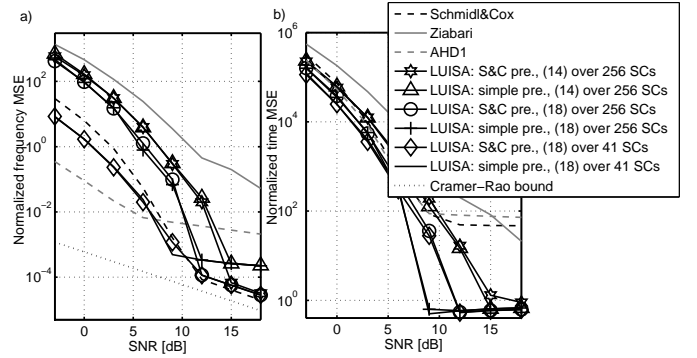}
\caption{Frequency (a) and time (b) MSE normalized to sampling period for all frames; no GSs, no interference. }
\label{fig_time_freq_MSE_no_interf_all}
\end{figure}
\begin{figure}[tbp]
\centering
\includegraphics[width=3.5in]{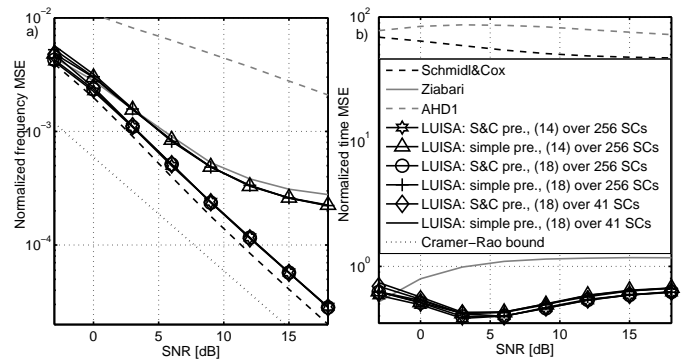}
\caption{Frequency (a) and time (b) normalized MSE for successfully synchronized frames; no GSs, no interference. }
\label{fig_time_freq_MSE_no_interf}
\end{figure}

In Fig. \ref{fig_time_freq_MSE_no_interf_all}, the MSEs are presented. The improved LUISA detection scheme, using (\ref{eq_improved_detection}), outperforms the simple detection using (\ref{eq_simple_detection}). Similarly, for low SNR, lower CFO search range decreases frequency-MSE and the simple preamble allows for lower time-MSE. While the CFO estimation in AHD1 is limited to the range of $(-2;2)$, the simulated CFO is in the range of $(-1.7;1.7)$.  It results in lower frequency-MSE obtained for low SNR. However, in the case of high SNR, the frequency-MSE floor is obtained of approximately $2\cdot 10^{-3}$ (as in \cite{Sanguinetti_2010_synchr}). Essentially, Ziabari method achieves both MSEs higher than the S\&C method although it has lower probability of synchronization error. 
This is because typically, an error for erroneously synchronized frame in S\&C algorithm is smaller than in the Ziabari method due to much wider range of integer CFO estimation in the latter. 
Therefore, it is reasonable to analyze MSEs only for correctly synchronized frames. In  
Fig. \ref{fig_time_freq_MSE_no_interf}, it can be observed that LUISA using the simple preamble, has frequency estimation error lower than both the Ziabari and AHD1 method. In the case of LUISA it reaches the MSE floor above $10^{-4}$. As explained previously, it is caused by non-zero variance of $Y_l(n,k)$. 
All LUISA configurations utilizing the S\&C preamble obtain similar frequency MSE which requires about 1~dB higher SNR than the S\&C algorithm.
The S\&C algorithm has theoretical MSE lower than LUISA (Sec. \ref{sec_fine_frac_CFO}) and all channel paths are inherently used for CFO estimation, that is not always true for LUISA.
The Cramer-Rao lower bound was defined in Sec. \ref{sec_fine_frac_CFO}. 
The plot in Fig. \ref{fig_time_freq_MSE_no_interf}.b shows time-MSE. For all LUISA configurations, the performance is similar, and typically this MSE is lower than in the competitive algorithms.
Interestingly, an increase of the number of propagation paths does not deteriorate LUISA performance (for a given $n$, all paths contribute to the variance of $Y(n,k)$, but maximally only one to its expected value). The simulation results (not shown here) reveal that even for 16 equal-mean-power-paths channel, LUISA keeps to outperform the other algorithms.

For the presence-of-interference scenario, we have chosen LUISA configuration showing the most of its advantages: the S\&C preamble, detection using formula (\ref{eq_improved_detection}) operating over 41 FFT outputs.  
\begin{figure}[tbp]
\centering
\includegraphics[width=3.5in]{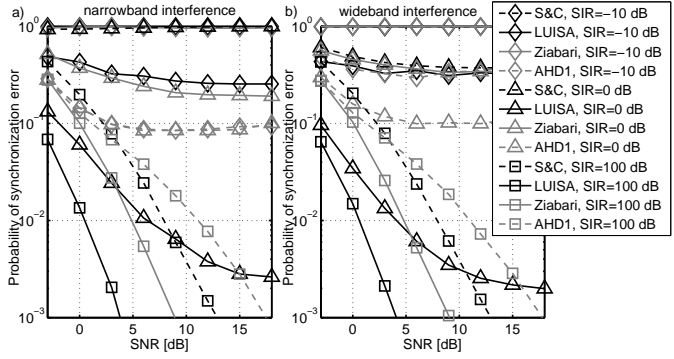}
\caption{Probability of synchronization error for an NC-OFDM system in the presence of NBI (a) and WBI (b); no GSs.}
\label{fig_prawd_narrow_notch}
\end{figure}
The probability of frame detection error is shown in Fig. \ref{fig_prawd_narrow_notch}. 
High probability of the synchronization error for S\&C algorithm with NBI (SIR$\leq$ 0 dB) is caused by the false synchronization effect. This is because the timing point is typically chosen in the inter-frame period, far from the optimal timing point. Thus, S\&C algorithm is not suitable even for coarse synchronization in the presence of high-power NBI.
The Ziabari method is not based on autocorrelation, and is not prone to this effect. In case of WBI, the  
Ziabari method is slightly better than the S\&C algorithm. These algorithms are outperformed by LUISA. The AHD1 is outperformed by LUISA in most cases, except for the NBI case and SIR=$-10$~dB. Below, we show that LUISA performance in the presence of NBI can be improved by GSs insertion or by integer-CFO search-range limitation.
For this case, MSE results are similar as in Fig. \ref{fig_time_freq_MSE_no_interf_all}.

\subsubsection{A system with GSs} 
\begin{figure}[!t]
\centering
\includegraphics[width=3.5in]{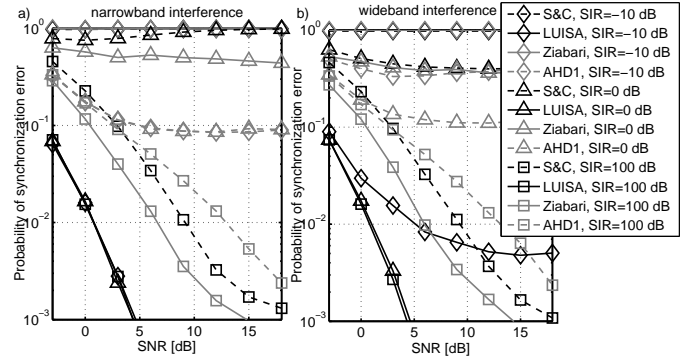}
\caption{Probability of synchronization error in the presence of NBI (a) and WBI (b); SSA scenario, GSs applied.}
\label{fig_prawd_wide_notch}
\end{figure}

Another test case assumes guard band of $15$ GSs on each side of the occupied band, i.e. $\mathbf{I}=\{-85,...,-1,1,47,...,85\}$. 
The example PSDs plots for the useful signal (denoted as NC-OFDM+GS) and both kinds of interference is shown in Fig. \ref{fig_widmo}. The resultant probabilities of synchronization error are shown in Fig. \ref{fig_prawd_wide_notch}. In comparison to the previous scenario, this probability is slightly increased for S\&C, AHD1 and Ziabari methods. Although the number of symbols modulating the preamble decreases, that makes the assumption of lack of correlation of time-domain samples (assumed in Sec. \ref{sec_statistical_properties}) less accurate, the performance of LUISA is improved, especially for low SIR values. 
The interference nearly does not increase the probability of synchronization error in comparison to the system with AWGN only. The perfect operation of LUISA with NBI is caused by the integer-CFO search-range limited to $k\in \{-20,...,20\}$ in $Y(n,k)$ and by the fact that the minimum distance in frequency between NBI and the occupied NC-OFDM band equals $47-24=23$. According to (\ref{eq_variance_accurate_interf_final}), this results in a zero variance of interference-based component in $Y(n,k)$ over the observed $k$ output range. In case of WBI, some interference-based variance adds to $Y(n,k)$ as shown in Fig. \ref{fig_variance_of_interf}. Thus, LUISA is highly suitable for NC-OFDM with LU signal occupying a different but adjacent band.   

\subsubsection{DSA scenario with GSs}
\begin{figure}[!t]
\centering
\includegraphics[width=3.5in]{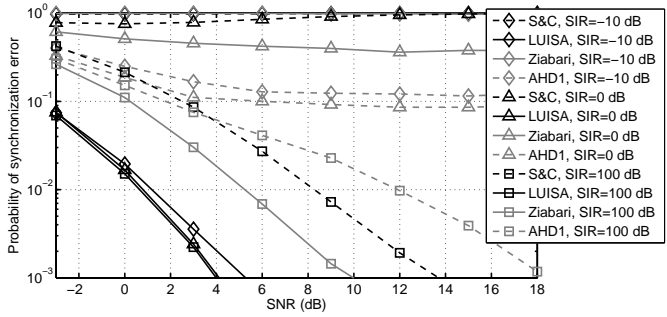}
\caption{Probability of synchronization error in the presence of NBI changing its center frequency for each frame; GSs applied.}
\label{fig_prawd_random_NBI}
\end{figure}

In this scenario, the frequency allocation of NC-OFDM and LU systems change dynamically and randomly and remain constant only within a single frame. Still, it requires the knowledge of the used SCs at the NC-OFDM receiver. The initial SCs set is $\mathbf{I}_{\mathrm{ini}}=\{-100,...,-1,1,...,100\}$. The NBI of random normalized frequency (of the uniform distribution over $(-128;127)$) is generated for each frame. This requires dynamic reconfiguration of NC-OFDM used SCs, 
and application of GSs. As above, the notch of $45$ SCs around NBI is created. The resultant probability of synchronization error is shown in Fig. \ref{fig_prawd_random_NBI}. Note that the performance of the S\&C, AHD1 and Ziabari method is close to this in Fig. \ref{fig_prawd_wide_notch}, and that LUISA outperforms all other schemes. A slight increase of LUISA synchronization-error probability for lower SIR is observed as a result of the non-integer NBI normalized frequency (some interference-related components of $\mathbb{V}\left[ Y(n,k)\right] $ is observed in the considered range of $k$ according to (\ref{eq_variance_accurate_interf_final})).

\subsection{Computational complexity}
In the considered algorithms, the operation dominating the complexity is the computation of the squared module of the synchronization variable ($|Y(n,k)|^{2}$ in case of LUISA). This complexity per sample is presented in Table \ref{table_computational_compl} for S\&C, Ziabari, AHD1 and LUISA algorithms.
In case of LUISA, we assume FFT using \emph{split-radix} method, preceded by $N$ complex multiplications. The calculations are presented for LUISA  
with increased robustness against the CFO using (\ref{eq_improved_detection}) over limited range of the integer CFO, 
where $\lceil\nu_{\mathrm{M}}\rceil=20$. It involves calculation of $Z(n,k)$ for $2\lceil\nu_{\mathrm{M}}\rceil$ inputs  followed by calculation of $|Y(n,k)|^{2}$ and $|Z(n,k)|^{2}$ over $2\lceil\nu_{\mathrm{M}}\rceil+1$  and $2\lceil\nu_{\mathrm{M}}\rceil$ elements, respectively. Note that LUISA has the complexity about 3 orders of magnitude higher than S\&C algorithm for $N=256$ (as in our simulations). However,
when comparing LUISA against more modern Ziabari and AHD1 algorithms, the increase in complexity is rather acceptable.
LUISA complexity can be further decreased by the optimization of FFT operations. Only $2\lceil\nu_{\mathrm{M}}\rceil+1$ $Y(n,k)$ output FFT samples are needed for a given $n$ and half of the complex multiplications, i.e., $N/2$, on FFT input can be carried inside FFT block for S\&C preamble. This can be done by the means of transform decomposition \cite{Sorensen_1993_FFT_pruning}. 
The approximate saving in real additions and multiplications is about 19\% and 8\% respectively in comparison to the basic LUISA. 
In some scenarios, utilization of (\ref{eq_Y_k_n_freq}) can be advantageous. It does not require FFT implementation because $R(n,k)$ can be calculated recursively using $R(n-1,k)$. In Table \ref{table_computational_compl}, the complexity-assessment results for this approach are also shown for the example assumption of 2 blocks of occupied SCs, S\&C preamble, no GSs, $\lceil\nu_{\mathrm{M}}\rceil\geq 1$ and QPSK modulation (multiplications are replaced by additions). The approach based on (\ref{eq_Y_k_n_freq}) is independent of $N$, and as such, can be a good choice for the scenario of low NC-OFDM subcarriers utilization, i.e. for $\alpha \ll N$. 

\linespread{1}
\begin{table}[thb]
\caption{Number of operations per single input sample $n$ for $N=256$}
\label{table_computational_compl}
\begin{center}
\begin{tabular}{lll}
\hline
Algorithm & Real additions/subtractions & Real multiplications \\ 
\hline
S\&C & $7$ & $6$ \\ 
Ziabari & $16N+11=4107$ & $16N+22=4118$ \\ 
AHD1 (iterative) & $12N+49=3121$ & $10\frac{3}{4}N+65=2817$ \\ 
LUISA & $3N\log_{2}N-N$ & $N\log_{2}N+N$ \\
 & $+12\lceil\nu_{\mathrm{M}}\rceil+5=6133$ & $+18\lceil\nu_{\mathrm{M}}\rceil+6=2670$\\
LUISA & $3N\log_{2}(2.8\lceil\nu_{\mathrm{M}}\rceil\!+\!4.2\!)$ & $N\log_{2}(2.8\lceil\nu_{\mathrm{M}}\!\rceil\! + \!4.2 \!)$ \\
(optimized, approx.) & $+N+8\lceil\nu_{\mathrm{M}}\rceil\!-\!1\!\approx \!4955$ & $+2N\!+\!22\lceil\nu_{\mathrm{M}}\rceil\!+\!2\!\approx\! 2467$\\
LUISA & $4\lceil\nu_{\mathrm{M}}\rceil \alpha+6\alpha$ & $4\alpha +34\lceil\nu_{\mathrm{M}}\rceil$ \\
(using (\ref{eq_Y_k_n_freq}), approx.) & $+28\lceil\nu_{\mathrm{M}}\rceil-7= 16463$ & $-14=1406$\\
\hline
\end{tabular}
\end{center}
\end{table}     
\linespread{1.55}

\section{Conclusions}
LUISA provides an SNR-gain in the probability of the synchronization error of a few dBs in comparison to the state-of-art algorithms. In the case of interference non-overlapping in frequency with NC-OFDM subcarriers, the gain is even higher. It is obtained at the cost of computational complexity increase over the S\&C method of 3 orders of magnitude. However, its computational complexity is comparable with Ziabari (10\% lower) or AHD1 (25\% higher) algorithms.


%

\appendices

\section{The mean and variance of $Y(n,k)$}
\label{sec_appendix_mean_and_var_Y_n_k}
We can calculate the expected value $\mathbb{E}[Y(n,k)]$ as follows\footnote{The proposed metric adds $N$ consecutive samples in time, that should keep the result close to the ensemble expectation.}:
\begin{align}
\!\!&\!\!\!\mathbb{E}[Y\!(n,k)]\!\!=\!\!\!\!\sum_{l=0}^{L-1}\!\!h(l)\mathbb{E}[Y_{l}\!(n,k)]
\!+\!\mathbb{E}[Y_{\mathrm{in}}\!(n,k)]\!\!+\!\!\mathbb{E}[Y_{\mathrm{no}}\!(n,k)].\!\!\!\!
\label{eq_exp_Y_k_n2}
\end{align}
The noise $w(n+m)$ is not correlated with the conjugate of the RP $x^{(0)*}_{m}$, so $\mathbb{E}[Y_{\mathrm{no}}(n,k)]=0$. Interference is also not correlated with the RP, and $\mathbb{E}[Y_{\mathrm{in}}(n,k)]=0$. It holds even in the case of NBI being the complex sinusoid of frequency $f_{\mathrm{NBI}}$, amplitude $A_{\mathrm{NBI}}$ and initial phase $\theta_{\mathrm{NBI}}$:
\begin{align}
\!\!\mathbb{E}[Y_{\mathrm{in}}(n,k)]\!\!=\!\!A_{\mathrm{NBI}} e^{j2\pi \frac{f_{\mathrm{NBI}} n}{N}\!+\!j\theta_{\mathrm{NBI}}}\!\!\!\!\sum_{m=0}^{N-1}\!\!\!
\mathbb{E}[x^{(0)*}_{m}]e^{\!\!-\!j2\pi \frac{(k-f_{\mathrm{NBI}}) m}{N}}\!\!=\!\!0.
\end{align}
The variance of $Y(n,k)$ is by definition:  $\mathbb{V}[Y(n,k)]=\mathbb{E}\left[Y(n,k)Y(n,k)^{*}\right]-\mathbb{E}\left[Y(n,k)\right]\mathbb{E}\left[Y(n,k)^{*}\right]$. As the second term is the squared absolute value of expectation defined in (\ref{eq_exp_Y_k_n_minimized}) and uses entries from the third column of Table \ref{table_variance_mean}, we will focus on the calculation of the first term. Let us substitute (\ref{eq_Y_k_n2}) to the first component of variance definition:  
\begin{align}
&
\mathbb{E}\left[Y\!(\!n,\!k)Y\!(\!n,\!k)^{*}\!\right]\!\!=\!\!\sum_{l_{1}=0}^{L-1}\!\sum_{l_{2}=0}^{L-1}\!h(l_{1})h(l_{2})^{*}\!\!\!
\sum_{m_{1}=0}^{N-1}\!\sum_{m_{2}=0}^{N-1}\!\! 
\nonumber \\&
\mathbb{E}\left[\tilde{x}(\!n\!+\!m_{1}\!\!-\!\!l_{1})x^{(0)*}_{m_{1}}\tilde{x}(n\!+\!m_{2}\!\!-\!\!l_{2})^{*}\!x^{(0)}_{m_{2}} \right] 
e^{j2\pi (m_{1}\!-\!m_{2})\frac{\nu-k}{N}}
\nonumber \\ &
\!\!+\!\!\!\sum_{l_{1}=0}^{L-1}\!\!h(l_{1})e^{j2\pi \frac{n\nu}{N}}\!\!\!\!\sum_{m_{1}=0}^{N-1}\!\sum_{m_{2}=0}^{N-1}
\!\!\!\mathbb{E}[\tilde{x}(n\!\!+\!\!m_{1}\!\!-\!\!l_{1})   
(i(n\!\!+\!\!m_{2})^{*}\!\!\!+\!\!w(n\!\!+\!\!m_{2})^{*}\!)
\nonumber \\&
\cdot
x^{(0)*}_{m_{1}}x^{(0)}_{m_{2}}]
e^{j\frac{2\pi}{N}(m_{1}(\nu-k)+k)} 
\!\!+\!\!\sum_{l_{2}=0}^{L-1}h(l_{2})^{*}e^{-j2\pi \frac{n\nu}{N}}\!\!
\nonumber \\ &
\!\!\sum_{m_{1}=0}^{N-1}\!\sum_{m_{2}=0}^{N-1}\!\!\mathbb{E}[\tilde{x}(n\!\!+\!\!m_{2}\!\!-\!\!l_{2})^{*} 
(i(n\!\!+\!\!m_{1})\!\!+\!\!w(n\!+\!m_{1}))x^{(0)}_{m_{1}}x^{(0)*}_{m_{2}}]
\nonumber \\ & 
e^{-j\frac{2\pi}{N}(m_{2}(\nu-k)+k)} 
\!\!+\!\!\sum_{m_{1}=0}^{N-1}\sum_{m_{2}=0}^{N-1}\mathbb{E}[(i(n+m_{1})\!+\!w(n+m_{1}))
\nonumber \\ &
\cdot
(i(n+m_{2})^{*}\!+\!w(n+m_{2})^{*})x^{(0)*}_{m_{1}}x^{(0)}_{m_{2}}]e^{j2\pi(m_{2}-m_{1})\frac{k}{N}}.\!
\label{eq_exp_Y_k_n_kwadrat4}
\end{align}
The first and the last component in (\ref{eq_exp_Y_k_n_kwadrat4}) can be non-zero while the other ones equal zero because the interference and noise are  not correlated with the NC-OFDM signal. Even for the NBI in the NC-OFDM spectrum notch and all three samples: $\tilde{x}(n+m_{2}-l_{2})$, $x^{(0)}_{m_{1}}$, $x^{(0)}_{m_{2}}$, aligned in time, non-central moments of the 3rd order of $x_{m}$,  $\mathbb{E}[x^{(0)*}_{m}x^{(0)}_{m}x^{(0)*}_{m}]$, equal 0. As the AWGN is not correlated with neither interference, nor with the RP, the last component (\ref{eq_exp_Y_k_n_kwadrat4}) can be simplified: 
\begin{align}
&\sum_{m_{1}=0}^{N-1}\sum_{m_{2}=0}^{N-1}\!\bigl(\mathbb{E}[
i(n\!+\!m_{1})i(\!n\!+\!m_{2}\!)^{*}x^{(0)*}_{m_{1}}x^{(0)}_{m_{2}}]\!
\nonumber \\ &
+\!\mathbb{E}[w(\!n\!+\!m_{1}\!)w(\!n\!+\!m_{2}\!)^{*}x^{(0)*}_{m_{1}}x^{(0)}_{m_{2}}]\bigr)
e^{j2\pi(m_{2}-m_{1})\frac{k}{N}}. 
\label{eq_exp_Y_k_n_kwadrat5}
\end{align}
The noise components are not correlated with the RP and are correlated with each other only for $m1=m2$, so $\mathbb{E}[w(\!n\!+\!m_{1}\!)w(\!n\!+\!m_{2}\!)^{*}x^{(0)*}_{m_{1}}x^{(0)}_{m_{2}}]
=\mathbb{E}[|w(\!n\!+\!m\!)|^{2}]\mathbb{E}[|x^{(0)}_{m}|^{2}]=\sigma^{2}_{\mathrm{w}}\sigma^{2}_{\mathrm{x}}$.
Similarly, it can be observed, that if $l_{1} \neq l_{2}$, the first component in (\ref{eq_exp_Y_k_n_kwadrat4}) equals zero, giving:
\begin{align}
\label{eq_exp_Y_k_n_kwadrat6}
&\mathbb{E}\left[Y(n,k)Y(n,k)^{*}\right]=N\sigma^{2}_{\mathrm{w}}\sigma^{2}_{\mathrm{x}}\!+\!\!
\sum_{l=0}^{L-1}|h(l)|^{2} 
\sum_{m_{1}=0}^{N-1}\sum_{m_{2}=0}^{N-1} 
 \\ &
\mathbb{E}\left[\tilde{x}(n+m_{1}-l)x^{(0)*}_{m_{1}}\tilde{x}(n+m_{2}-l)^{*}x^{(0)}_{m_{2}} \right]
e^{j2\pi (m_{1}-m_{2})\frac{\nu-k}{N}}
\nonumber \\&
+\!\!\sum_{m_{1}=0}^{N-1}\sum_{m_{2}=0}^{N-1}\!\mathbb{E}[
i(n\!+\!m_{1})i\!(\!n\!+\!m_{2}\!)^{*}x^{(0)*}_{m_{1}}x^{(0)}_{m_{2}}]
e^{j2\pi(m_{2}-m_{1})\frac{k}{N}}.
\nonumber
\end{align}
Assuming interference model (\ref{eq_interf_freq}) and (\ref{eq_IFFT}) for RP ($p=0$), the interference-involving component in (\ref{eq_exp_Y_k_n_kwadrat6}) can be determined using summation of geometric progressions over $m_1$ and $m_{2}$:  
\begin{align}
&\!\!\!\!\!\!\sum_{m_{1}=0}^{N-1}\sum_{m_{2}=0}^{N-1}\!\mathbb{E}[
i(n\!+\!m_{1})i\!(\!n\!+\!m_{2}\!)^{*}x^{(0)*}_{m_{1}}x^{(0)}_{m_{2}}]
e^{j2\pi(m_{2}-m_{1})\frac{k}{N}}
\!
\nonumber \\ &
\!\!=\!\!
\frac{1}{N^{2}}\!\!\!\!
\sum_{k1=-N/2}^{N/2-1}\!\sum_{k2=-N/2}^{N/2-1}\!\sum_{k3=-N/2}^{N/2-1}
\!\sum_{k4=-N/2}^{N/2-1}\!\!\!\!\!
\mathbb{E}[g_{k1}\!(\!n\!)g_{k2}\!(\!n\!)^{*}d^{(0)*}_{k3}d^{(0)}_{k4}]
\nonumber \\&
\cdot
e^{j\pi \frac{(N-1)(k1\!-\!k2\!-\!k3)}{N}}
\frac{\sin\left(\pi (k1\!-\!k3\!-\!k)\right)}
{\sin\left(\pi \frac{k1-k3-k}{N}\right)}
\frac
{\sin\left(\pi (k4-k2+k)\right)}
{\sin\left(\pi \frac{k4-k2+k}{N}\right)}
\nonumber \\ &
=
\sum_{k_{1}=0}^{N-1}\sum_{k_{2}=0}^{N-1}\!\mathbb{E}[ 
g_{k1}(n)d^{(0)*}_{(k1-k)^{\circ}}g^{(n)*}_{k2}d^{(0)}_{(k2-k)^{\circ}}
].\!
\label{eq_exp_Y_k_n_kwadrat_accurate_interf}
\end{align}
Observe that both periodic sinc functions in (\ref{eq_exp_Y_k_n_kwadrat_accurate_interf}) can be non-zero only for
$k1-k3-k\in\{-N,0,N\}$ and $k4-k2+k\in\{-N,0,N\}$, respectively (because $k,k1,k2, k3, k4\in\{-N/2,...,N/2-1\}$).
Note that summation over expectations in the first component in (\ref{eq_exp_Y_k_n_kwadrat6}) can be done for each path independently and that preamble symbols are correlated only when $k1=k2$ in (\ref{eq_exp_Y_k_n_kwadrat_accurate_interf}) that allows for simple formula for the variance of $Y(n,k)$:
\begin{align}
\mathbb{V}\left[Y(n,k)\right]=&\sum_{l=0}^{L-1}|h(l)|^{2}\mathbb{V}\left[ Y_{l}(n,k)\right]
\label{eq_variance_accurate_interf_final}
 \\&\nonumber
+\sum_{k_{1}=0}^{N-1}
\mathbb{E}[ 
|g_{k1}(n)|^{2}]\mathbb{E}[|d^{(0)}_{(k1-k)^{\circ}}|^{2}]
+N\sigma^{2}_{\mathrm{w}}\sigma^{2}_{\mathrm{x}}.
\end{align}
If there is no correlation between samples $\tilde{x}(n+m_{1}-l)$ and $x^{(0)*}_{m_{1}}$, i.e., when the received signal is not aligned with any of possible cases A--D from Fig. \ref{fig_corr_cases} $\mathbb{E}[\tilde{x}(n+m1-l)x^{(0)*}_{m1}]=0$, and:
\begin{align}
&\mathbb{V}\!\left[Y_{l}(n\!,\!k)\right]\!\!=\!\!\!\!\!\sum_{m_{1}=0}^{N-1}\sum_{m_{2}=0}^{N-1}\!\!\!\mathbb{E}[x^{(0)*}_{m_{1}}x^{(0)}_{m_{2}}]  
\mathbb{E}[\tilde{x}(n\!+\!m_{1}\!-\!l)\tilde{x}(n\!\!+\!\!m_{2}\!\!-\!l)^{*}]
\nonumber \\ &
\cdot e^{j2\pi (m_{1}\!-\!m_{2})\frac{\nu-k}{N}}\!\!\!-\!\!|\mathbb{E}[Y_{l}(n,k)]|^{2}\!\! 
=\!\!\!\sum_{m=0}^{N-1}\!\!\mathbb{E}[x^{(0)*}_{m}x^{(0)}_{m}]
\mathbb{E}[\tilde{x}(\!n\!+\!m\!-\!l\!)
\nonumber \\&
\cdot\tilde{x}(\!n\!\!+\!\!m\!\!-\!\!l\!)^{*}]\!-\!0
\!=\!\!\sigma^{2}_{\mathrm{x}}\sigma^{2}_{\mathrm{x}}\!\min(\max(0,\!N\!\!+\!\!N_{\mathrm{CP}}\!\!+\!\!n\!\!-\!\!l\!),N).
\label{eq_variance_no_corr}
\end{align}

\section{The mean value of $Y_{l}(l,k)$ for case C and the S\&C preamble}
\label{sec_appendix_mean}
The mean value of (\ref{eq_exp_Y_k_n_l}) in case C can be rewritten as
\begin{align}
&\mathbb{E}[Y_{l}(l,k)]=e^{j2\pi \frac{l\nu}{N} } \sum_{m=0}^{N-1}\mathbb{E}[x^{(0)}_{m}x^{(0)*}_{m}]e^{j2\pi\frac{m(\nu-k)}{N}}.
\label{eq_exp_Y_k_n_l_appendix}
\end{align}
In the specific case when preamble consists of two identical halves
\begin{align}
\!\!\!\!\!\!\mathbb{E}[Y_{l}(l,k)]\!\!=&e^{j2\pi \frac{l\nu}{N} } \!\!\!\sum_{m=0}^{N/2-1}\!\!\!\!\left(\!1\!+\!e^{j\pi (\nu-k)}\right)\!\mathbb{E}[x^{(0)}_{m}x^{(0)*}_{m}]
e^{j2\pi\frac{m(\nu-k)}{N}}\!\!
\nonumber \\&
=\!e^{j2\pi \frac{l\nu}{N}}\!\left(\!1\!+\!e^{j\pi (\nu-k)}\!\right)\!\sigma^{2}_{\mathrm{x}}\!\!\!\sum_{m=0}^{N/2-1}\!\!e^{j2\pi\frac{m(\nu-k)}{N}}.\!\!
\label{eq_exp_Y_k_n_l_appendix2}
\end{align}
By using the formula for the sum of a geometric sequence and further simplifications we get
\begin{align}
&\mathbb{E}[Y_{l}(l,k)]\!=\!e^{j2\pi \frac{l\nu}{N}}\!\left(\!1\!+\!e^{j\pi (\nu-k)}\!\right)\!\sigma^{2}_{\mathrm{x}}\!\frac{1-e^{j2\pi\frac{N}{2}\frac{\nu-k}{N}}}{1-e^{j2\pi\frac{\nu-k}{N}}}
\nonumber \\ &
=e^{j2\pi l\frac{\nu}{N}}\sigma^{2}_{\mathrm{x}}e^{j\pi (\nu-k)\left(1-\frac{1}{N}\right)}\frac{\sin\left(\pi (\nu-k) \right)}{\sin\left(\pi \frac{\nu-k}{N} \right)}.
\label{eq_exp_Y_k_n_l_appendix3}
\end{align}
\section{The variance of $Y_{l}(l,k)$ for case C and the S\&C preamble}
\label{sec_appendix_variance}
In Case C, $n=l$. Thus, the variance of a single-path useful signal component in (\ref{eq_variance_accurate_interf_final}) is calculated for $Y_{l}(l,k)$ instead of $Y_{l}(n,k)$ giving:
\begin{align}
\mathbb{V}[Y_{l}(l,k)]=&\sum_{m_{1}=0}^{N-1}\sum_{m_{2}=0}^{N-1}
\mathbb{E}\left[\left|x^{(0)}_{m_{1}}\right|^{2}\left|x^{(0)}_{m_{2}}\right|^{2}\right]e^{j2\pi (m_{1}-m_{2})\frac{\nu-k}{N}}
\nonumber\\&
-\left|\mathbb{E}\left[Y_{l}(l,k)\right]\right|^{2}. \label{eq_VAR_C_1}
\end{align}
Both parts of the preamble are identical (i.e. $N/2$ samples) that makes us to convert both summations (over $m_{1}$ and $m_{2}$) to the range from $0$ to $N/2-1$, that results in:  
\begin{align}
&\mathbb{V}[Y_{l}(l,k)]=\sum_{m_{1}=0}^{N/2-1}\sum_{m_{2}=0}^{N/2-1}
\mathbb{E}\left[\left|x^{(0)}_{m_{1}}\right|^{2}\left|x^{(0)}_{m_{2}}\right|^{2}\right]
\nonumber\\&
\!\left[e^{j2\pi (m_{1}-m_{2})\frac{\nu-k}{N}}\!+\!e^{j2\pi (m_{1}\!-\!m_{2}\!+\!\frac{N}{2})\frac{\nu-k}{N}}\!+\!e^{j2\pi (m_{1}\!+\!\frac{N}{2}\!-\!m_{2}\!-\!\frac{N}{2})\frac{\nu-k}{N}}\right.
\nonumber\\&
\left.
+e^{j2\pi (m_{1}-m_{2}-\frac{N}{2})\frac{\nu-k}{N}}\right]
-\left|\mathbb{E}\left[Y_{l}(l,k)\right]\right|^{2}.\label{eq_VAR_C_2}
\end{align}
Common part of exponential components can be placed in front of the summation, that can be divided into two cases, i.e. $m_{1}=m_{2}$ and $m_{1}\neq m_{2}$,  giving
\begin{align}
&\mathbb{V}[Y_{l}(l,k)]=\Big(2+2\cos \left(\pi (\nu-k)\right)\Big)\bigg(
\sum_{m_{1}=0}^{N/2-1} \mathbb{E}\left[\left|x^{(0)}_{m_{1}}\right|^{4}\right] \nonumber\\&
+\!\!\!\sum_{m_{1}=0}^{N/2-1}\sum_{m_{2}=0, m_{2}\neq m_{1}}^{N/2-1}
\!\!\!\!\mathbb{E}\left[\left|x^{(0)}_{m_{1}}\right|^{2}\right] \!\mathbb{E}\left[\left|x^{(0)}_{m_{2}}\right|^{2}\right]
\!e^{j2\pi (m_{1}-m_{2})\frac{\nu-k}{N}}\!\bigg)
\nonumber \\&
-\left|\mathbb{E}\left[Y_{l}(l,k)\right]\right|^{2}\!\!=\!\!
\Big(2\!+\!2\cos \left(\pi (\nu\!-\!k)\right)\Big)\Bigg(\Bigg.
\!\sum_{m_{1}=0}^{N/2-1}\!\! 2 \sigma^{4}_{\mathrm{x}}\!-\!\sigma^{4}_{\mathrm{x}}\frac{N}{2}
 \nonumber \\&
\left.+\sum_{m_{1}=0}^{N/2-1}\sum_{m_{2}=0}^{N/2-1}
\sigma^{4}_{\mathrm{x}}e^{j2\pi (m_{1}-m_{2})\frac{\nu-k}{N}}\right)-\left|\mathbb{E}\left[Y_{l}(l,k)\right]\right|^{2},
\label{eq_VAR_C_3}
\end{align}
as each $x^{(0)}_{m_{1}}$ composes of independent real and imaginary parts, each of variance $\sigma^{2}_{\mathrm{x}}/2$. Thus, $\mathbb{E}\left[\left|x^{(0)}_{m_{1}}\right|^{4}\right]=2\sigma^{4}_{\mathrm{x}}$. The second summation can be decomposed into two separate sums (over $m_{1}$ and $m_{2}$) over geometric sequences giving:
\begin{align}
&\mathbb{V}[Y_{l}(l,k)]\!=\!\Big(\!2\!+\!2\cos \left(\pi (\nu\!-\!k)\right)\!\!\Big)\!
\Bigg(\!\!\sigma^{4}_{\mathrm{x}}\frac{N}{2}\label{eq_VAR_C_4}
\nonumber\\&
\!+\!\frac{\left(1-e^{\pi(\nu-k)}\right)\left(1-e^{-\pi(\nu-k)}\right)}{\left(1-e^{2\pi\frac{\nu-k}{N}}\right)\left(1-e^{-2\pi\frac{\nu-k}{N}}\right)}\Bigg)
-\!\left|\mathbb{E}\left[Y_{l}(l,k)\right]\right|^{2}.
\end{align}
After substitution of $\mathbb{E}\left[Y_{l}(l,k)\right]$ from Table \ref{table_variance_mean} we get:
\begin{align}
&\mathbb{V}[Y_{l}(l,k)]\!=\!\Big(2\!+\!2\cos \left(\pi (\nu\!-\!k)\right)\!\!\Big)\!
\Bigg(\sigma^{4}_{\mathrm{x}}\frac{N}{2}
\nonumber \\&
\!+\!\sigma^{4}_{\mathrm{x}} \left(\frac{\sin(\frac{\pi}{2}(\nu-k))}{\sin(\frac{\pi}{N}(\nu-k))}\!\right)^{2}\!\Bigg)
\!\!-\!\!\sigma^{4}_{\mathrm{x}} \left(\!\frac{\sin(\pi(\nu-k))}{\sin(\frac{\pi}{N}(\nu-k))}\!\right)^{2}\!\!,
\label{eq_VAR_C_5}
\end{align}
that can be further simplified using trigonometrical identities
\begin{align}
&\mathbb{V}[Y_{l}(l,k)]=\Big(2+2\cos \left(\pi (\nu-k)\right)\Big)
\sigma^{4}_{\mathrm{x}}\frac{N}{2}
\label{eq_VAR_C_6}
\\&
+\sigma^{4}_{\mathrm{x}}\left(\frac{\sin(\frac{\pi}{2}(\nu-k))}{\sin(\frac{\pi}{N}(\nu-k))}\right)^{2}
\Bigg(\Big( 2+2\cos \left(\pi (\nu-k)\right) \Big)\nonumber \\
&-4 \left(\cos\left(\frac{\pi}{2}(\nu-k)\right) \right)^2 \Bigg)
=\bigg(2+2\cos\left(\pi(\nu-k)\right)\bigg)\sigma^{4}_{\mathrm{x}}\frac{N}{2}.
\nonumber
\end{align}

\ifCLASSOPTIONcaptionsoff
  \newpage
\fi




\bibliographystyle{IEEEtran}
\bibliography{pawla_bib}

\begin{thebibliography}{10}
\providecommand{\url}[1]{#1}
\csname url@samestyle\endcsname
\providecommand{\newblock}{\relax}
\providecommand{\bibinfo}[2]{#2}
\providecommand{\BIBentrySTDinterwordspacing}{\spaceskip=0pt\relax}
\providecommand{\BIBentryALTinterwordstretchfactor}{4}
\providecommand{\BIBentryALTinterwordspacing}{\spaceskip=\fontdimen2\font plus
\BIBentryALTinterwordstretchfactor\fontdimen3\font minus
  \fontdimen4\font\relax}
\providecommand{\BIBforeignlanguage}[2]{{%
\expandafter\ifx\csname l@#1\endcsname\relax
\typeout{** WARNING: IEEEtran.bst: No hyphenation pattern has been}%
\typeout{** loaded for the language `#1'. Using the pattern for}%
\typeout{** the default language instead.}%
\else
\language=\csname l@#1\endcsname
\fi
#2}}
\providecommand{\BIBdecl}{\relax}
\BIBdecl

\bibitem{Bogucka_Radar_Sonar_2011}
S.~Pagadarai, A.~Kliks, H.~Bogucka, and A.~Wyglinski, ``Non-contiguous
  multicarrier waveforms in practical opportunistic wireless systems,''
  \emph{Radar, Sonar Navigation, IET}, vol.~5, no.~6, pp. 674--680, July 2011.

\bibitem{Mahmoud09_OFDM_for_CR}
H.~Mahmoud, T.~Yucek, and H.~Arslan, ``{OFDM} for cognitive radio: merits and
  challenges,'' \emph{Wireless Communications, IEEE}, vol.~16, no.~2, pp.
  6--15, 2009.

\bibitem{Weiss_oknowanie_Re_i_TX}
T.~Weiss, J.~Hillenbrand, A.~Krohn, and F.~Jondral, ``Mutual interference in
  {OFDM}-based spectrum pooling systems,'' in \emph{Vehicular Technology
  Conference, 2004. VTC 2004-Spring. 2004 IEEE 59th}, vol.~4, 2004, pp.
  1873--1877 Vol.4.

\bibitem{KryszkiewiczEURASIP12}
P.~Kryszkiewicz, H.~Bogucka, and A.~Wyglinski, ``Protection of primary users in
  dynamically varying radio environment: practical solutions and challenges,''
  \emph{EURASIP Journal on Wireless Communications and Networking}, vol. 2012,
  no.~1, p.~23, 2012.

\bibitem{Kryszkiewicz_OCCS_2013}
P.~Kryszkiewicz and H.~Bogucka, ``Out-of-band power reduction in {NC-OFDM} with
  optimized cancellation carriers selection,'' \emph{Communications Letters,
  IEEE}, vol.~PP, no.~99, pp. 1--4, 2013.

\bibitem{Brandes_mitigation_NB_interf}
S.~Brandes, M.~Schnell, U.~Berthold, and F.~Jondral, ``{OFDM} based overlay
  systems - design challenges and solutions,'' in \emph{IEEE PIMRC 2007}, 2007,
  pp. 1--5.

\bibitem{Morelli_07_overview_synchr}
M.~Morelli, C.-C. Kuo, and M.-O. Pun, ``Synchronization techniques for
  orthogonal frequency division multiple access ({OFDMA}): A tutorial review,''
  \emph{Proceedings of the IEEE}, vol.~95, no.~7, pp. 1394--1427, 2007.

\bibitem{Schmidl_Cox_1997}
T.~Schmidl and D.~Cox, ``Robust frequency and timing synchronization for
  {OFDM},'' \emph{Communications, IEEE Transactions on}, vol.~45, no.~12, pp.
  1613--1621, 1997.

\bibitem{Minn_2003_improved_S_C}
H.~Minn, V.~Bhargava, and K.~Letaief, ``A robust timing and frequency
  synchronization for {OFDM} systems,'' \emph{Wireless Communications, IEEE
  Transactions on}, vol.~2, no.~4, pp. 822--839, 2003.

\bibitem{Morelli_freq_improv_S_C}
M.~Morelli and U.~Mengali, ``An improved frequency offset estimator for {OFDM}
  applications,'' \emph{Communications Letters, IEEE}, vol.~3, no.~3, pp.
  75--77, 1999.

\bibitem{Coulson_2004_synchr_narrow_sinusoid}
A.~Coulson, ``Narrowband interference in pilot symbol assisted {OFDM}
  systems,'' \emph{Wireless Communications, IEEE Transactions on}, vol.~3,
  no.~6, pp. 2277--2287, 2004.

\bibitem{Marey_2007_SC_narrowband}
M.~Marey and H.~Steendam, ``Analysis of the narrowband interference effect on
  {OFDM} timing synchronization,'' \emph{Signal Processing, IEEE Transactions
  on}, vol.~55, no.~9, pp. 4558--4566, 2007.

\bibitem{Zivkovic_2011_synchr_SC_wideband}
M.~Zivkovic and R.~Mathar, ``Performance evaluation of timing synchronization
  in {OFDM}-based cognitive radio systems,'' in \emph{Vehicular Technology
  Conference (VTC Fall), 2011 IEEE}, 2011, pp. 1--5.

\bibitem{Awoseyila_synchronization}
A.~Awoseyila, C.~Kasparis, and B.~Evans, ``Robust time-domain timing and
  frequency synchronization for {OFDM} systems,'' \emph{Consumer Electronics,
  IEEE Transactions on}, vol.~55, no.~2, pp. 391--399, 2009.

\bibitem{Sun_2010_synchr_filter}
P.~Sun and L.~Zhang, ``Timing synchronization for {OFDM} based spectrum sharing
  system,'' in \emph{Wireless Communication Systems (ISWCS), 2010 7th
  International Symposium on}, 2010, pp. 951--955.

\bibitem{Faulkner_filtering_effect_OFDM}
M.~Faulkner, ``The effect of filtering on the performance of {OFDM} systems,''
  \emph{Vehicular Technology, IEEE Transactions on}, vol.~49, no.~5, pp.
  1877--1884, 2000.

\bibitem{Sanguinetti_2010_synchr}
L.~Sanguinetti, M.~Morelli, and {\relax H.V}.~Poor, ``Frame detection and
  timing acquisition for {OFDM} transmissions with unknown interference,''
  \emph{Wireless Communications, IEEE Transactions on}, vol.~9, no.~3, pp.
  1226--1236, March 2010.

\bibitem{Saha_Dyspan_synchronization}
D.~Saha, A.~Dutta, D.~Grunwald, and D.~Sicker, ``Blind synchronization for
  {NC-OFDM} when channels are conventions, not mandates,'' in \emph{IEEE DySPAN
  2011}, 2011, pp. 552--563.

\bibitem{Ziabari_synchr_NC_OFDM}
H.~Abdzadeh-Ziabari and M.~Shayesteh, ``Robust timing and frequency
  synchronization for {OFDM} systems,'' \emph{Vehicular Technology, IEEE
  Transactions on}, vol.~60, no.~8, pp. 3646--3656, 2011.

\bibitem{Ren_freq_estimation}
G.~Ren, Y.~Chang, H.~Zhang, and H.~Zhang, ``An efficient frequency offset
  estimation method with a large range for wireless {OFDM} systems,''
  \emph{Vehicular Technology, IEEE Transactions on}, vol.~56, no.~4, pp.
  1892--1895, 2007.

\bibitem{Oberg_2001_modulation_detection_coding}
T.~Oberg, \emph{Modulation, Detection and Coding}.\hskip 1em plus 0.5em minus
  0.4em\relax New York, NY, USA: John Wiley \& Sons, Inc., 2001.

\bibitem{Wei_2010_rozklad_OFDM}
S.~Wei, D.~Goeckel, and P.~Kelly, ``Convergence of the complex envelope of
  bandlimited {OFDM} signals,'' \emph{Information Theory, IEEE Transactions
  on}, vol.~56, no.~10, pp. 4893--4904, Oct 2010.

\bibitem{Yucek07}
T.~Yucek and H.~Arslan, ``{MMSE} noise plus interference power estimation in
  adaptive {OFDM} systems,'' \emph{Vehicular Technology, IEEE Transactions on},
  vol.~56, no.~6, pp. 3857--3863, Nov 2007.

\bibitem{Candan_freq_est2013}
C.~Candan, ``Analysis and further improvement of fine resolution frequency
  estimation method from three {DFT} samples,'' \emph{Signal Processing
  Letters, IEEE}, vol.~20, no.~9, pp. 913--916, Sept 2013.

\bibitem{Ozdemir_2007_OFDM_channel_est}
M.~Ozdemir and H.~Arslan, ``Channel estimation for wireless ofdm systems,''
  \emph{Communications Surveys Tutorials, IEEE}, vol.~9, no.~2, pp. 18--48,
  Second 2007.

\bibitem{3gpp.36.101}
3GPP, ``{Evolved Universal Terrestrial Radio Access (E-UTRA); User Equipment
  (UE) radio transmission and reception},'' {3rd Generation Partnership Project
  (3GPP)}, TS {36.101}, Sep. 2008.

\bibitem{Sorensen_1993_FFT_pruning}
H.~Sorensen and C.~Burrus, ``Efficient computation of the {DFT} with only a
  subset of input or output points,'' \emph{Signal Processing, IEEE
  Transactions on}, vol.~41, no.~3, pp. 1184--1200, Mar 1993.

\end{thebibliography}
%

%







\end{document}